\documentclass[aps,twocolumn,floatfix]{revtex4}
\usepackage{amsmath,amssymb}
\usepackage{bm}
\usepackage{graphicx}
\usepackage{subfigure}
\usepackage{psfrag}

\newcommand{\figwidth}{6 cm}
\newcommand{\identity}{\mathit{I}}
\newcommand{\one}{\mathbb{I}}
\newcommand{\ket}[1]{\left| \:#1\: \right>}
\newcommand{\bra}[1]{\left< \:#1\: \right|}
\newcommand{\braket}[2]{\left< \:#1\: \right. | \left. \:#2\: \right>}

\newcommand{\sgn}{{\rm sgn}}

\begin{document}
\title{
Excitations from Filled Landau Levels in Graphene
}
\author{A. Iyengar$^1$, Jianhui Wang$^1$, H. A. Fertig$^1$, L. Brey$^2$}
\affiliation{$^1$Department of Physics, Indiana University, Bloomington, IN  47405}
\affiliation{$^2$Instituto de Ciencia de Materiales de Madrid (CSIC), Catoblanco, 28049 Madrid, Spain}
\date{\today}

\begin{abstract}
We consider graphene in a strong perpendicular magnetic field at zero temperature
with an integral number of filled Landau levels 
and study the dispersion of single particle-hole excitations.
We first analyze the two-body problem of a single Dirac electron and hole in a magnetic 
field interacting via Coulomb forces.  
We then turn to the many-body problem, where particle-hole symmetry and the 
existence of two valleys lead to a number of effects peculiar to graphene.
We find that the coupling together of a large number of low-lying excitations leads to
strong many-body corrections, which could be observed in inelastic light scattering or optical
absorption.  We also discuss in detail how the appearance of different branches in the exciton dispersion
is sensitive to the number of filled spin and valley sublevels.
\end{abstract}

\maketitle

\section{Introduction}

Recent experimental progress has allowed the fabrication of
{\it graphene}, a two-dimensional honeycomb lattice of carbon
atoms that form the basic planar structure in graphite \cite{graphene}.  Graphene
has a host of interesting properties descending from its unusual
band structure, which includes linear spectra around two inequivalent
points in the Brillouin zone.  The eigenfunctions and eigenvalues
of the single-particle electron quantum states near these points
are well-described by the Dirac equation \cite{review,brey1,brey2}, 
where the wavefunctions have a spinor structure due to the
two-point basis of the honeycomb lattice. The wavefunctions
in the vicinity of each of the Dirac points are chiral,
leading to a Berry's phase when a particle is dragged around
either one of them.  This has important consequences for
transport, most notably an absence of backscattering that
inhibits localization \cite{ando98} and a finite conductivity
even when undoped \cite{zhang}.
In the presence of a magnetic field, the graphene structure has been
theoretically shown to shift both Shubnikov-de Haas oscillations \cite{miktik99},
and the step pattern of the integer quantized Hall effect \cite{zheng}.
Both these effects have recently been confirmed in experiment \cite{zhang}.

All these properties may be understood in terms of non-interacting
electrons in the graphene bandstructure.  How do electron-electron
interactions impact the properties of graphene?  To sensibly explore
this question requires an understanding of the low-lying excitations
of the interacting system.  Various parts of these spectra may
in some cases be measured by inelastic light scattering \cite{pinczuk}
or microwave absorption \cite{Sadowski}.
In this work, we
present a study of these excitations in the quantum Hall regime.  

For a standard
2DEG, a seminal paper by Kallin and Halperin \cite{KallHalp}
demonstrated that for densities such that the non-interacting spectrum --
highly degenerate Landau levels with a harmonic oscillator spectrum --
has its Fermi energy in a gap, the low-lying collective mode 
spectrum may be interpreted
in terms of a single particle-hole pair excited across the gap.
The attractive interaction between these particles binds them into
an exciton, which carries a well-defined momentum $\bm{P}$ in spite
of the magnetic field because the object as a whole is neutral.
At small $P$, the exciton dispersion behaves as some form of
density wave, typically either a magnetoplasmon or a spin density
wave, depending on the quantum numbers of the electron and hole.
At larger $P$, the dispersion crosses over smoothly into a form
expected for a particle-hole pair  with separation 
$P \ell^2$, with $\ell=\sqrt{\hbar c/eB}$ the magnetic length
and $B$ the magnetic field.  A simple classical interpretation
of the excitation in this limit is that the mutual attraction
of the particle and hole at some separation precisely cancels
the Lorentz force when they move together at an appropriate
velocity, establishing a connection between the momentum of the
object and the particle-hole separation.

In graphene this simple picture necessarily breaks down, because in the
absence of a magnetic field, particles and holes move at a constant
speed set by the linear spectrum around the Dirac points.  Nevertheless,
as we shall see the results for the quantum spectrum of these excitations
are qualitatively quite similar to those of the 2DEG.  The main differences
arise from the relatively large number of states within each Landau level --
there are two spin states and two valley states for each Landau state --
which allows for sixteen different excitations when the electron and
hole have different Landau indices.  We shall further see that the
large number of levels leads to stronger many-body corrections to
the simple two-body model described above than is the case for the 2DEG.

In what follows we will examine the exciton spectrum of this system
in two different ways, first as a two-body problem of a positively
and negatively charged particle each with a graphene spectrum in a magnetic
field, and then as a many-body problem in which a single particle is 
excited across the Fermi level of a filled Fermi sea.  At large momentum
$P$ we will see these spectra largely agree (up to a constant which
is undetermined in the first approach), while at small $P$ there can be
considerable difference.

This article is organized as follows: 
Section \ref{sec:two-body} begins with a discussion of non-interacting
Dirac particles in a magnetic field.  After reviewing the spectrum and wavefunctions, we turn 
to the two-body problem, where we introduce a canonical transformation to separate the 
relative motion. We solve for the spectrum in the presence of the Coulomb interaction, 
incorporating several Landau levels with an eye towards measuring the degree of Landau-level mixing. 
In Section \ref{sec:many-body}, we introduce the many-body problem with a detailed discussion of the
low-lying excitations of the non-interacting system, for a variety of configurations of 
the chemical potential.  The mixing of some of these excitations can be anticipated from the symmetries
of the Hamiltonian.  The momentum-dependent exciton energies are then calculated
and catalogued according to spin and pseudospin quantum numbers.  We conclude by highlighting some salient
features of the neutral excitations in graphene, contrasting them with those of the 2DEG.  
The calculation of matrix elements and dispersions is detailed in the Appendix.

\section{Two-body problem}
\label{sec:two-body}

The continuum limit of non-interacting graphene is obtained by adopting a $\bm{k} \cdot \bm{p}$ approximation
\cite{review} about each of the two Dirac points, which we label using a pseudospin
$t_z$ with eigenvalues $\pm 1/2$, denoted by $\Uparrow$ and $\Downarrow$.
For a given spin state, the sublattice ($A, B$) and pseudospin ($\Uparrow , \Downarrow$) 
degrees of freedom lead to a four-component wavefunction, 
which we write in the basis $(\Uparrow \! A, \Uparrow \! B, \Downarrow \! A, \Downarrow \! B)$.
In addition to the Zeeman splitting, we allow for a much smaller pseudospin splitting $\hbar \omega_t$.
The one-body Hamiltonian can be written $H_0 = H_{\rm kin} - |g \mu_B B| \hat{s}_z - \hbar \omega_t \hat{t}_z$.
In the absence of a magnetic field, $H_{\rm kin}$ is given in our four-component basis by 
\begin{eqnarray}
\label{eq:Hkin}
\hbar v_F \: \left(
\begin{array}{cccc}
0 & p_x + i p_y & 0 & 0 \\
p_x - i p_y & 0 & 0 & 0 \\
0 & 0 & 0 & p_x - i p_y \\
0 & 0 & p_x + i p_y & 0  
\end{array}
\right) . 
\end{eqnarray}
In a magnetic field, we substitute $\bm{p} \rightarrow \bm{p} - \bm{A}$.
In the Landau gauge, $\bm{A} = x \hat{y} / \ell^2$, and $p_y$ becomes a good quantum number.
For the most part we
will use the units $\ell = \hbar = v_F = 1$, showing these constants explicitly when appropriate.
We define a dimensionless constant $\beta \equiv (e^2 / \epsilon \ell)/(\hbar v_F / \ell)$ to 
measure the approximate ratio of kinetic and potential energies.
In these units, the Coulomb potential $u(\bm{r}) \equiv e^2/(\epsilon r)$ becomes $\beta / r$.

For an eigenstate with $p_y = k$, we define the annihilation operator
$c = (p_x - i x + i k)/\sqrt{2}$, and in a magnetic field we have 
\begin{eqnarray}
H_{\rm kin} = \sqrt{2} \left(
\begin{array}{cccc}
0 & c & 0 & 0 \\
c^{\dagger} & 0 & 0 & 0 \\
0 & 0 & 0 & c^{\dagger} \\
0 & 0 & c & 0 \\
\end{array}
\right) , 
\end{eqnarray}
which describes each valley as a Dirac oscillator.
The eigenstates of such oscillators are enumerated by 
an integer Landau level index which can take \emph{any} value, 
positive, negative, or zero.  We employ the condensed notation for single-particle eigenstates 
$1 = (n_1,k_1,t_1,s_1)$ with energy 
$\varepsilon_1$, where the respective labels designate the  
Landau level, $p_y$, pseudospin, and spin of the state $1$. 
For a given spin state,
the eigenfunctions in the Landau level $n$ corresponding 
to $t_z = +1/2$, $t_z = -1/2$ are, respectively, 
\begin{eqnarray}
L_y^{-1/2} e^{i k y} & & \frac{1}{\sqrt{2}}\left( 
\begin{array}{c}
\sgn(n) \varphi_{|n| - 1}(x - k) \\
\varphi_{|n|}(x - k) \\
0 \\ 0 
\end{array}
\right) \: , 
\\ \nonumber 
L_y^{-1/2} e^{i k y} & & \frac{1}{\sqrt{2}}\left( 
\begin{array}{c}
0 \\ 0 \\ 
\varphi_{|n|}(x - k) \\
\sgn(n) \varphi_{|n| - 1}(x - k) 
\end{array}
\right)
\end{eqnarray}
for $n \ne 0$ and
\begin{eqnarray}
L_y^{-1/2} e^{i k y} & & \left( 
\begin{array}{c}
0 \\
\varphi_{0}(x - k) \\
0 \\ 0 
\end{array}
\right) \: , 
\\ \nonumber 
L_y^{-1/2} e^{i k y} & & \left( 
\begin{array}{c}
0 \\ 0 \\ 
\varphi_{0}(x - k) \\
0
\end{array}
\right)
\end{eqnarray}
for $n = 0$,
where $L_y$ is the length of the system in the $y$ direction, and 
$\varphi_n(x)$ are the usual wavefunctions of a simple harmonic oscillator, 
given in the Appendix. These states have energy 
\begin{eqnarray}
\varepsilon = \sqrt{2} (\hbar v_F / \ell) \sgn(n)|n|^{1/2} - |g \mu_B B| s_z - \hbar \omega_t \:t_z, 
\end{eqnarray}
showing the $\sqrt{B} |n|^{1/2}$ energy dependence of the Landau levels and the  
particle-hole symmetry characteristic of graphene.

We will frequently use a more compact notation, representing wavefunctions in 
component form as
\begin{eqnarray}
[\psi_{n,k}^t(x,y)]_{\tau} &\equiv& L_y^{-1/2} e^{i k y} 
[\sqrt{2}]^{\delta_{n,0}-1} 
\\ & & \nonumber 
s(n,t,\tau) \varphi_{\lambda(n,t,\tau)}(x - k) , 
\end{eqnarray}
where $\tau$ labels the sublattice, and
\begin{eqnarray}
s(n,t,\tau) &=& \left\{
\begin{array}{lcc}
\sgn(n) & ; & t = \Uparrow, \tau = A \\
 & \:{\rm or}\: & t = \Downarrow , \tau = B \\
1 & ; & t = \Uparrow, \tau = B \\
& \:{\rm or}\: & t = \Downarrow , \tau = A 
\end{array}
\right. \\
\lambda(n,t,\tau) &=& \left\{
\begin{array}{lcc}
|n| - 1 & ; & t = \Uparrow, \tau = A \\
 & \:{\rm or}\: & t = \Downarrow , \tau = B \\
|n| & ; & t = \Uparrow, \tau = B \\
& \:{\rm or}\: & t = \Downarrow , \tau = A 
\end{array}
\right. 
.
\end{eqnarray}

For the two-body problem, we neglect Zeeman and pseudospin splitting and consider only the kinetic
energy of the electron (1) and hole (2), described by the Hamiltonian
\begin{eqnarray}
H = H_{\rm kin}^{(1)} (\bm{A}) 
+ H_{\rm kin}^{(2)} (- \bm{A}) - u(\bm{r}^{(1)} - \bm{r}^{(2)}),  
\end{eqnarray}
where the sign of the vector potential is reversed in the hole term to account for its
opposite charge.  Without loss of generality, we may assume that the electron and hole are 
in the $\Uparrow$ valley. One may obtain other cases
by interchanging the sublattices of one or both particles, which does not change the 
spectrum since the kinetic energy is degenerate in the two valleys and 
the Coulomb interaction does not scatter between valley states in our 
continuum approximation \cite{CoulombApprox}. 
With the pseudospin fixed, we need only consider the upper left block of $H_{\rm kin}$ in Eq. 
\ref{eq:Hkin}, which can be conveniently expressed using the Pauli matrices $\sigma_i$ and the 
$2 \times 2$ identity matrix $\identity$.  On the other hand, $4 \times 4$ matrices are required to 
describe the action of the Hamiltonian on the sublattice degrees of freedom of the two-body system, 
and we express these matrices using a tensor product of two $2 \times 2$ matrices 
in which the left and right slots correspond to the action on the 
electron and hole spinors, respectively.
We define center-of-mass and relative coordinates as 
$\bm{p}^{(1)} = \bm{P}/2 + \bm{p}$,
$\bm{p}^{(2)} = \bm{P}/2 - \bm{p}$,
$\bm{r}^{(1)} = \bm{R} + \bm{r}/2$,
$\bm{r}^{(2)} = \bm{R} - \bm{r}/2$.
We now have
\begin{eqnarray}
H &=& P_x m^+_x + (P_y - x) m^+_y 
\\ & & \quad + \nonumber 2 p_x m^-_x + 2(p_y - X) m^-_y - u(\bm{r}) \: \one
\end{eqnarray}
where 
$m^{\pm}_x \equiv \frac{1}{2}(\sigma_x \otimes \identity \pm \identity \otimes \sigma_x)$, 
$m^{\pm}_y \equiv - \frac{1}{2}(\sigma_y \otimes \identity \pm \identity \otimes \sigma_y)$,
and $\one \equiv \identity \otimes \identity$.
To decouple the center-of-mass motion, we introduce the transformation
$U = e^{i X y}$, for which 
$U^{\dagger} P_x U = P_x + y$ and
$U^{\dagger} p_y U = p_y + X$.
The result is 
\begin{align}
U^{\dagger}H U = 
2 \bm{p} \cdot \bm{m}^- + (\bm{P} - \hat{z} \times \bm{r}) \cdot \bm{m}^+
- u(r) \: \one .
\end{align}
It is now apparent that $\bm{P}$ commutes with the transformed Hamiltonian
and can be replaced by a center-of-mass momentum quantum number.
For a given $\bm{P}$, we may study the interacting problem in the relative coordinate
since $U^{\dagger} u(r) U = u(r)$.
The operator $U \bm{P} U^{\dagger}$  commuting with the original Hamiltonian 
is identical to the two-body momentum operator $\bm{Q}$ found in 
Ref. \onlinecite{KallHalp}. 
When $\bm{P} = 0$, the system possesses an additional
symmetry consisting of a simultaneous relative spatial coordinate and SU(2) sublattice rotation.
We will return to this later. 

We may now shift $\bm{r} \rightarrow \bm{r} - \hat{z} \times \bm{P}$, moving the $\bm{P}$-dependence 
to the potential energy. The kinetic energy is now 
$2 \bm{p} \cdot \bm{m}^- - (\hat{z} \times \bm{r}) \cdot \bm{m}^+$, 
which describes a two-dimensional Dirac oscillator. To see this, define the 
harmonic lowering operators $a = p_x - ix/2$, $b = p_y - iy/2$.  We can also 
take the combination $c_{\pm} = (a \pm i b)/\sqrt{2}$, which describes independent
``+'' and ``-'' oscillators.
Using the four-component basis 
$(A^{(1)} A^{(2)},
A^{(1)} B^{(2)},
B^{(1)} A^{(2)},
B^{(1)} B^{(2)})$ 
for the sublattices of the electron and hole, 
\begin{eqnarray}
2 \bm{p} \cdot \bm{m}^- - (\hat{z} \times \bm{r}) \cdot \bm{m}^+ =
\sqrt{2} \left(
\begin{array}{cccc}
0 & c_-^{\dagger} & c_+ & 0 \\
c_- & 0 & 0 & c_+ \\
c_+^{\dagger} & 0 & 0 & c_-^{\dagger} \\
0 & c_+^{\dagger} & c_- & 0
\end{array}
\right),
\end{eqnarray}
or alternatively, in our tensor product notation,
\begin{eqnarray}
\sqrt{2} \left[ 
\left(
\begin{array}{cc}
0 & c_+ \\
c_+^{\dagger} & 0
\end{array}
\right) \otimes \identity
+ 
\identity \otimes \left(
\begin{array}{cc}
0 & c_-^{\dagger} \\
c_- & 0
\end{array}
\right) 
\right] .
\end{eqnarray}
It is now clear that the two-dimensional Dirac oscillator is characterized by \emph{two} quantum numbers
$n_+$ and $n_-$, which are the Landau level index of the electron and hole, respectively. 

The eigenstates of the kinetic energy can be written in terms of the 
two-dimensional harmonic oscillator wavefunctions $\Phi_{n_1,n_2}(\bm{r})$, defined
in the Appendix.  For an electron in Landau level $n_+$ and hole in level $n_-$, 
the four-component wavefunctions for the relative coordinate are 
\begin{eqnarray}
[\sqrt{2}]^{\delta_{n_+,0} + \delta_{n_-,0} - 2} \left(
\begin{array}{c}
s_+ s_- \Phi_{|n_+| - 1,|n_-| - 1}(\bm{r}) \\
s_+ \Phi_{|n_+| - 1,|n_-|}(\bm{r}) \\
s_- \Phi_{|n_+|,|n_-| - 1}(\bm{r}) \\
\Phi_{|n_+|,|n_-|}(\bm{r}) 
\end{array}
\right) 
\end{eqnarray}
where $s_{\pm} = \sgn(n_{\pm})$.
The corresponding kinetic energy is 
$\sqrt{2}\:[\sgn(n_+)|n_+|^{1/2} - \sgn(n_-)|n_-|^{1/2}]$.

We pause to consider the virtues of the canonical transformation. 
It maps the space of momentum states of an electron and hole, each of which carries
a one-dimensional momentum in our gauge, to a two-dimensional space of momentum states labelled by $\bm{P}$.
The relative 
motion is described by a two-dimensional Dirac oscillator moving in a $\bm{P}$-dependent
potential, whose two non-interacting quantum numbers are the Landau level indices of the electron and hole.
It is worth noting that the separation of the center-of-mass coordinates relies 
on the uniformity of the magnetic field and on the overall neutrality of the system.  When these 
conditions are met, such a transformation is always possible, although it will vary
depending on the choice of gauge.
It has an essential two-body nature and cannot be reduced to a choice of single-particle basis. 

Furthermore, the transformation is not peculiar to graphene 
and is equally applicable to the 2DEG, where one obtains a two-dimensional \emph{harmonic} oscillator. 
In this case, the operators $c_+$ and $c_-$ operate on independent clockwise and counter-clockwise modes.
As a result, the angular momentum of the oscillator corresponds to the difference between the
Landau level indices of the electron and hole; indeed, $\Phi_{n_+,n_-}(\bm{r})$ has 
azimuthal dependence $e^{- i \phi (n_+ - n_-)}$, where $\phi$ is the azimuthal angle of $\bm{r}$.  
This motivates the nomenclature $l_z \equiv n_+ - n_-$ for the difference between the Landau level indices 
of the electron and hole.
When $P=0$, the effective potential for the oscillator $u(r - \hat{z} \times \bm{P})$ becomes
rotationally symmetric, and matrix elements between two-body states with different values 
of $l_z$ will vanish.  The conservation of $l_z$ at $P=0$ can be viewed as a conservation
law for the internal angular momentum of the exciton.

An analogous conservation law arises in graphene.  
However, $l_z$ is not conserved, since in an eigenstate of the kinetic energy, the 
different spinor components can have different azimuthal dependences.  
We find instead that when $\bm{P} = 0$, the quantity $|n_+| - |n_-|$ is conserved by interactions.
Physically, this is due to the aforementioned $\bm{P} = 0$ symmetry involving combined coordinate and sublattice
rotations. This is the graphene analog of $l_z$ conservation in the 2DEG.

To summarize, the dispersion of the two-body problem is obtained by diagonalizing the relative coordinate
Hamiltonian, whose matrix elements $h$ are given by
\begin{align}
&h_{n_+, n_- ; n_+', n_-'}(\bm{P}) = 
\\
& \nonumber \quad \delta_{n_+,n_+'} \delta_{n_-,n_-'} \sqrt{2} 
(s_+ |n_+|^{1/2}
- s_- |n_-|^{1/2})
\\ & \nonumber \quad - [\sqrt{2}]^{
\delta_{n_+,0} + 
\delta_{n_-,0} + 
\delta_{n_+',0} + 
\delta_{n_-',0}
-4} \cdot [
\\ & \nonumber \quad \quad 
s_+ s_- s_+' s_-' u_{
|n_+| - 1, |n_-| - 1; 
|n_+'| - 1, |n_-'| - 1
}
\\ &  \nonumber \quad \quad +  
s_+ s_+' u_{
|n_+| - 1, |n_-|; 
|n_+'| - 1, |n_-'|
}
\\ &  \nonumber \quad \quad +  
s_- s_-' u_{
|n_+|, |n_-| - 1; 
|n_+'|, |n_-'| - 1
} 
\\ &  \nonumber \quad \quad +
u_{
|n_+|, |n_-|; 
|n_+'|, |n_-'|
} ]
\end{align}
where
\begin{align}
\label{eq:uformula}
u_{n_1,n_2;n_3,n_4}(\bm{P}) &\equiv \\ \nonumber 
\int d \bm{r} \: &
u(\bm{r} - \hat{z} \times \bm{P})
\Phi^*_{n_3,n_4}(\bm{r})
\Phi_{n_1,n_2}(\bm{r}) .
\end{align}

The location of the chemical potential will determine the possible Landau level indices for the electron and hole.
The splitting of each level into 4 spin and pseudospin sublevels enlarges the realm of possibilities. 
We use the notation $\gamma_n$ for the number of filled sublevels in Landau level $n$, where $n$
refers to the highest Landau level with filled sublevels. 
Thus, the undoped system with chemical potential $\mu = 0$ corresponds to $\gamma_0 = 2$, whereas placing 
the chemical potential between Landau levels $n$ and $n+1$ is denoted as $\gamma_n = 4$.
We consider the two-body problem both
for $\gamma_n = 4$ and $\gamma_n < 4$, neglecting the contribution of Zeeman and 
valley splitting to the non-interacting energy.
The cases $\gamma_n = 1,2,3$, which are essentially identical in this two-body analysis,  
raise the possibility that both the particle and hole are in level $n$, giving 
rise to intra- Landau level excitons carrying spin or pseudospin. 

Notably, the typical 
ratio of kinetic and potential energies $\beta$ is of order unity and independent of the 
strength of the magnetic field.  
We would like to establish whether Landau level mixing is weak in this situation, 
particularly in order to justify approximations used in the many-body treatment of the next section.
We have $\beta \equiv (e^2 / \epsilon \ell)/(\hbar v_F / \ell) = \alpha (c/v_F \epsilon)$, where 
$\alpha \approx 1/137$ is the fine-structure constant.
Experiments both in transport \cite{graphene,zhang} and far-infrared absorption 
\cite{Sadowski} find $c / v_F \approx 300$. 
Alicea and Fisher \cite{Alicea} estimate $\epsilon = 5$ for graphene grown on a
SiO$_2$ substrate, which accounts both for screening currents in the substrate as well 
as weak intrinsic screening in graphene computed in the RPA\cite{GonzalezScreening}.  
The screening should be somewhat better on a SiC substrate. We assume the more 
pessimistic value $\epsilon = 3$, leading to $\beta = 1/1.37 \approx 0.73$.

We study the cases $\gamma_n = 4$ and $\gamma_n < 4$ for $n=0,1,2,3$, computing the dispersion of the 
first five two-body states up to $P\ell = 12$ using 4 electron and 4 hole levels.
The ket $\ket{n_+, n_-}$ denotes a non-interacting state with the electron in level $n_+$ 
and the hole in level $n_-$.
We will see that, even though $\beta$ is of order unity, 
Landau level mixing plays a small role for these excitations.
Figure \ref{fig:dispersion} shows the case $\gamma_n = 4$.
The dispersions approach the non-interacting energies of the states $\ket{n_+,n_-}$, shown as dotted lines,
as $P \rightarrow \infty$. An analysis of Eq. \ref{eq:uformula} shows that the limit is approached as $1/P$.
For $n=0$, the non-interacting degeneracy in the third excited state (between $\ket{4,0}$ and $\ket{1,-1}$) 
is lifted at low $P$ by interactions. 
The dispersions computed from a single state $\ket{n_+,n_-}$, without Landau level mixing, are 
shown as dashed lines.  
Although $\beta \sim 1$, the effect of mixing is relatively weak for many of the exciton states.
The first and second excited states for $n=2$ and $n=3$ are notable exceptions.
This can be understood by noting that the 
non-interacting states $\ket{n+1, n-1}$ and $\ket{n+2,n}$ can admix strongly 
at small $P$ since they have the 
same nearly-conserved quantum number $|n_+| - |n_-|$.
For the ground state, the main qualitative effect of Landau level mixing 
in all four cases seems to be the deepening of peaks and dips in the dispersion. 

The same observations apply to the case $\gamma_n < 4$, shown in Figure \ref{fig:dispersion_gne4}.
Here, there are non-interacting degeneracies in the excited states for $n=0$, and in the third 
excited state for $n=1$.  Note that in the former case, this degeneracy is lifted at low $P$ only 
after the inclusion of Landau level mixing.  The degeneracy is not broken for $P=0$, at which 
$|n_+| - |n_-|$ is exactly conserved.  
Due to the small (possibly zero) kinetic energies, 
some exciton energies become negative.  
This situation is rectified in the many-body calculation of the next section
when self-energy effects are included.

\begin{psfrags}
\psfrag{Energy}[Bc][Bc]{Energy $\: (\hbar v_F / \ell)$}
\psfrag{Pl}{$P \ell$}
\psfrag{40}{$\ket{4,0}$,}
\psfrag{30}{$\ket{3,0}$}
\psfrag{9}{$\:\ket{1,-1}$}
\psfrag{20}{$\ket{2,0}$}
\psfrag{100}{$\ket{1,0}$}
\psfrag{51}{$\ket{5,1}$}
\psfrag{41}{$\ket{4,1}$}
\psfrag{31}{$\ket{3,1}$}
\psfrag{21}{$\ket{2,1}$}
\psfrag{11}{$\ket{1,1}$}
\psfrag{52}{$\ket{5,2}$}
\psfrag{42}{$\ket{4,2}$}
\psfrag{32}{$\ket{3,2}$}
\psfrag{22}{$\ket{2,2}$}
\psfrag{63}{$\ket{6,3}$}
\psfrag{53}{$\ket{5,3}$}
\psfrag{43}{$\ket{4,3}$}
\psfrag{33}{$\ket{3,3}$}
\psfrag{02201}{$\ket{0,-2}$,}
\psfrag{02202}{$\:\ket{2,0}$}
\psfrag{01101}{$\ket{0,-1}$,}
\psfrag{01102}{$\:\ket{1,0}$}
\psfrag{00}{$\ket{0,0}$}
\psfrag{10411}{$\ket{1,0}$,}
\psfrag{10412}{$\:\ket{4,1}$}
\begin{figure*}
\centering
\subfigure[$\gamma_0 = 4$]{
\includegraphics[width=\figwidth]{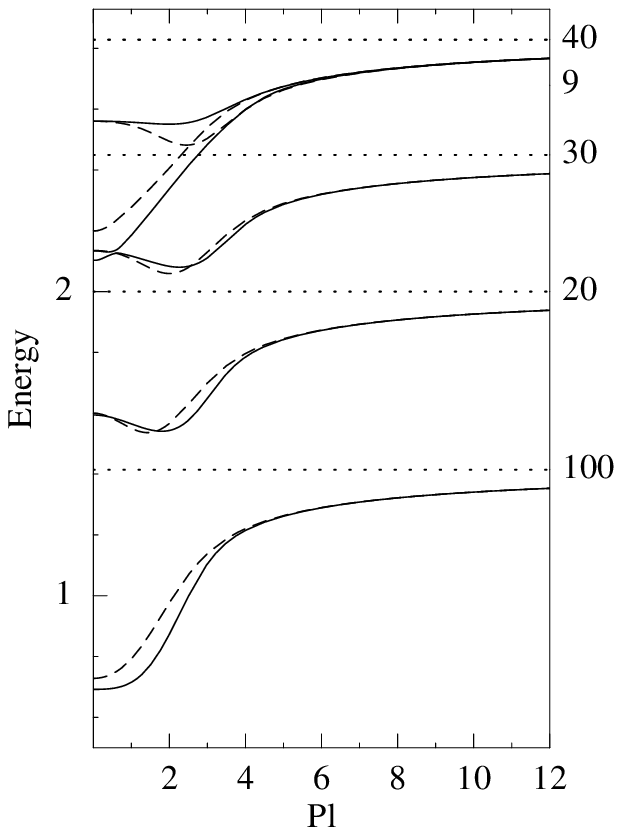}
}
\qquad 
\subfigure[$\gamma_1 = 4$]{
\includegraphics[width=\figwidth]{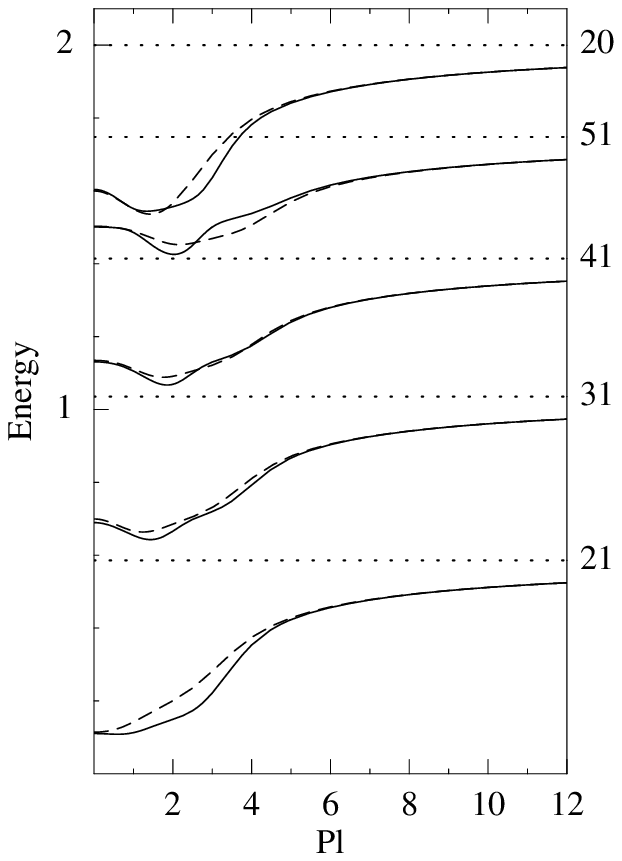}
} \\
\subfigure[$\gamma_2 = 4$]{
\includegraphics[width=\figwidth]{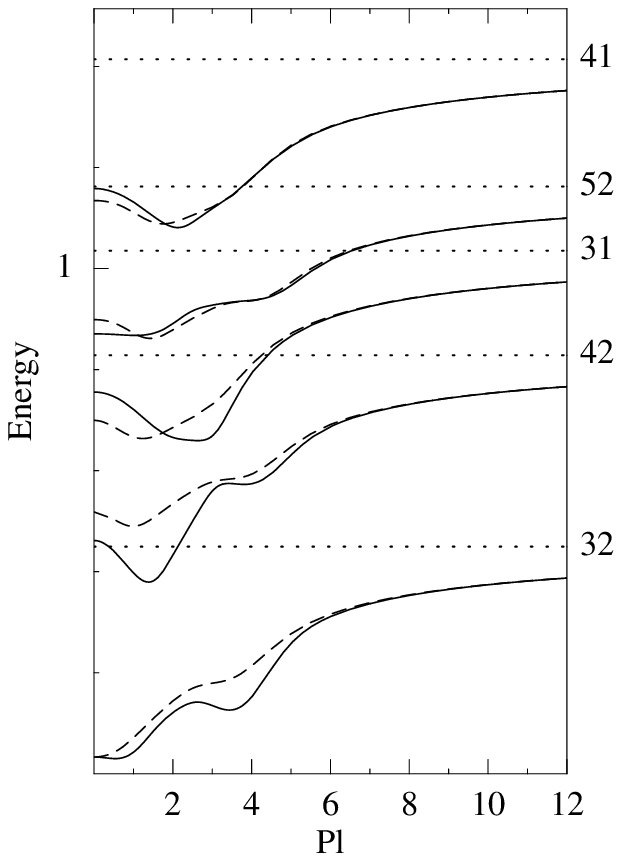}
}\qquad
\subfigure[$\gamma_3 = 4$]{
\includegraphics[width=\figwidth]{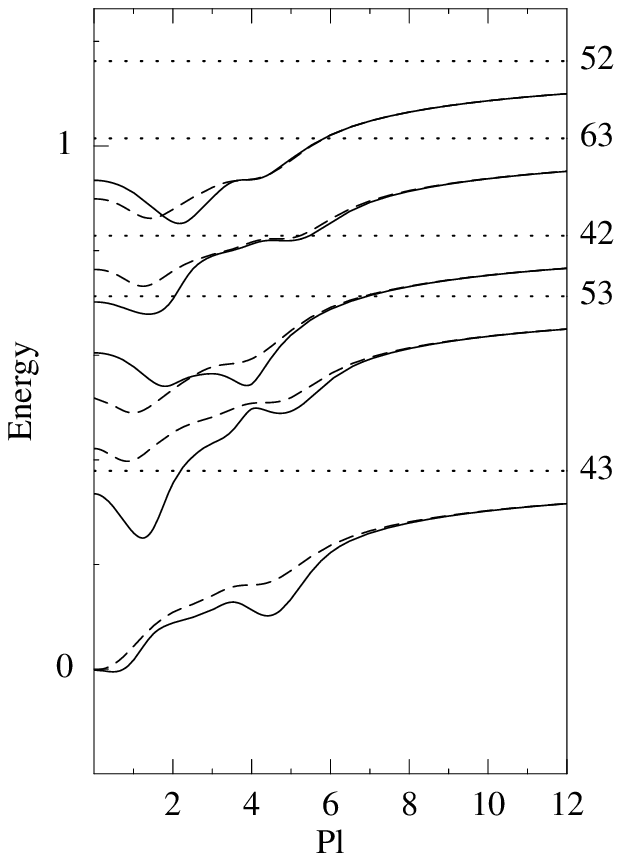}
} 
\caption{Dispersion of the first few levels for $\beta = 0.73$
when $\gamma_n = 4$. Dashed curves neglect Landau level mixing.
}
\label{fig:dispersion}
\end{figure*}

\begin{figure*}
\centering
\subfigure[$\gamma_0 < 4$]{
\includegraphics[width=\figwidth]{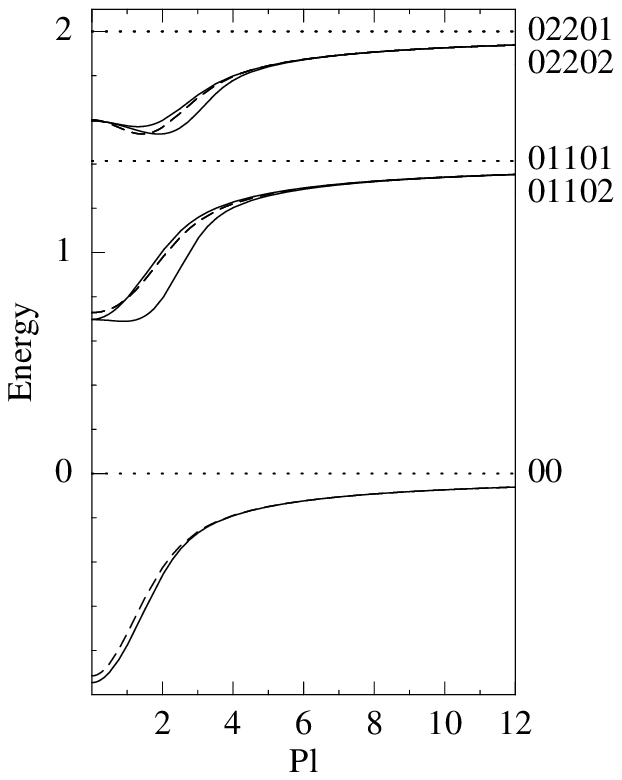}
}
\qquad 
\subfigure[$\gamma_1 < 4$]{
\includegraphics[width=\figwidth]{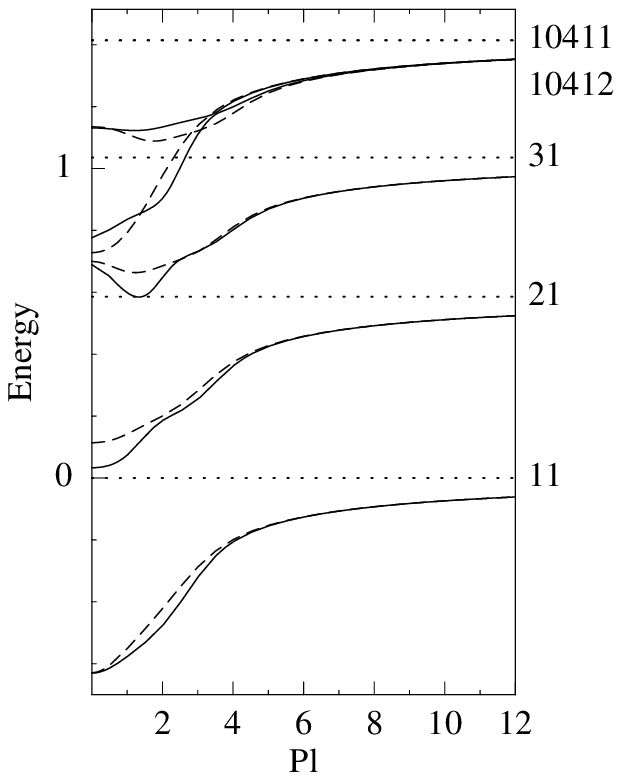}
} \\
\subfigure[$\gamma_2 < 4$]{
\includegraphics[width=\figwidth]{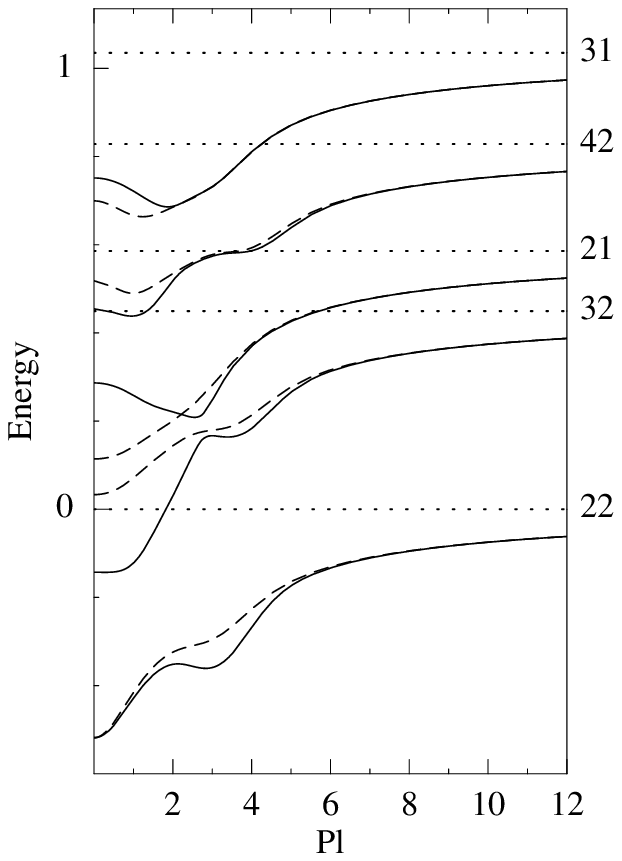}
}\qquad
\subfigure[$\gamma_3 < 4$]{
\includegraphics[width=\figwidth]{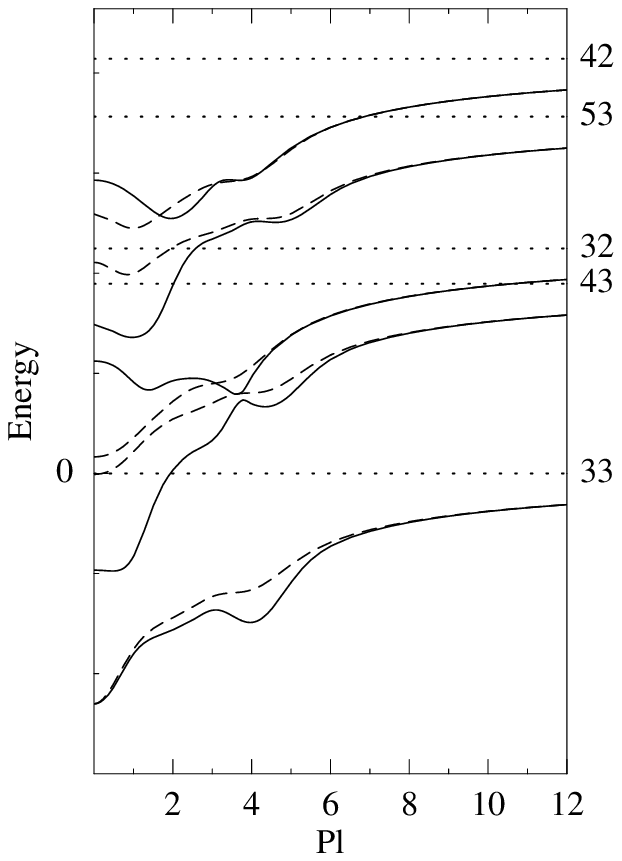}
} 
\caption{Dispersion of the first few levels for $\beta = 0.73$
when $\gamma_n < 4$. Dashed curves neglect Landau level mixing.
}
\label{fig:dispersion_gne4}
\end{figure*}
\end{psfrags}

\begin{figure}
\psfrag{Pl}{$P \ell$}
\includegraphics[width=\columnwidth]{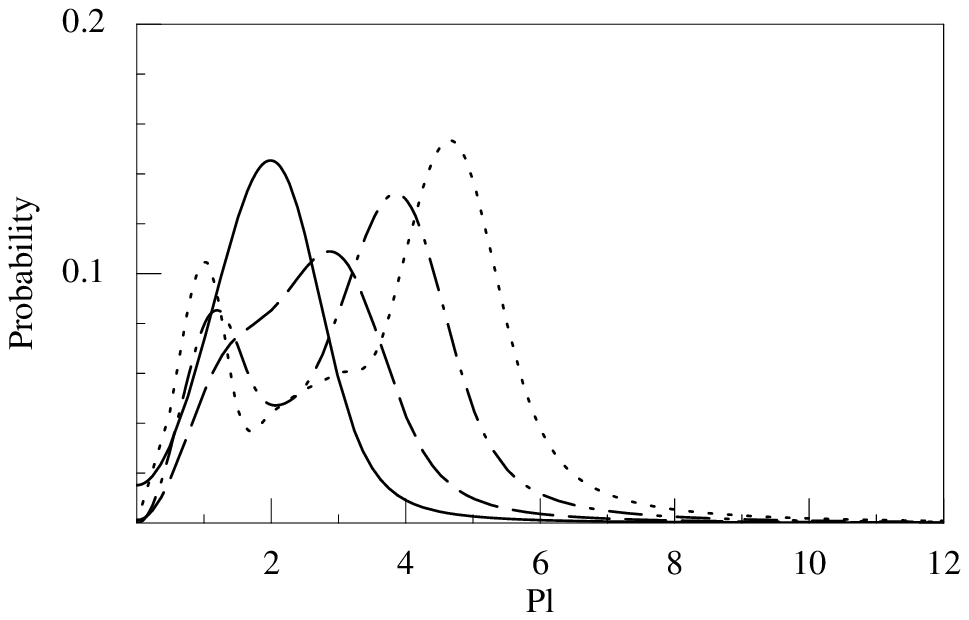}
\caption{
Amount of Landau level mixing at momentum $P$, as defined by Eq. \ref{eq:mixing}, for $\beta = 0.73$.
Results shown for 
$\gamma_0 = 4$ (solid), 
$\gamma_1 = 4$ (dashed),
$\gamma_2 = 4$ (dot-dashed),
$\gamma_3 = 4$ (dotted).
}
\label{fig:prob_nu}
\end{figure}

\begin{figure}
\psfrag{Pl}{$P \ell$}
\includegraphics[width=\columnwidth]{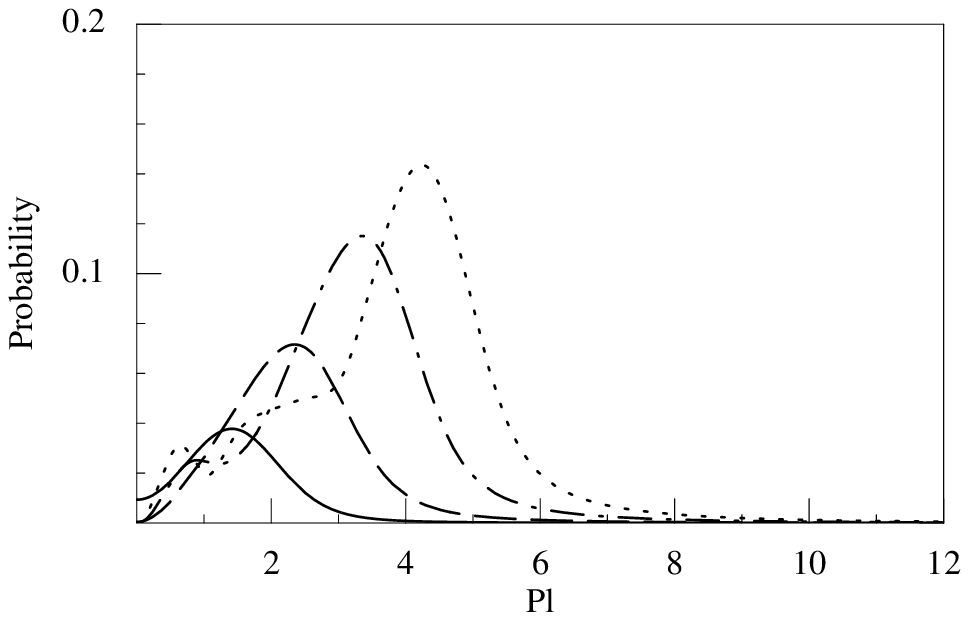}
\caption{
Amount of Landau level mixing at momentum $P$, as defined by Eq. \ref{eq:mixing}, for $\beta = 0.73$.
Results shown for 
$\gamma_0 < 4$ (solid), 
$\gamma_1 < 4$ (dashed),
$\gamma_2 < 4$ (dot-dashed),
$\gamma_3 < 4$ (dotted).
}
\label{fig:prob_gne4}
\end{figure}

We pursue Landau level mixing further in Figures \ref{fig:prob_nu} and \ref{fig:prob_gne4}, which show the 
square-modulus of the overlap between the ground state (denoted $\ket{0}$) and higher non-interacting states, given by
\begin{eqnarray}
\begin{array}{ccl}
1 - |\braket{0}{n + 1, n}|^2 &;& \gamma_n = 4 \\
1 - |\braket{0}{n, n}|^2 &;& \gamma_n < 4 .
\end{array}
\label{eq:mixing}
\end{eqnarray}
The level of mixing is 15\% or less for all cases, vanishing rapidly in the limit $P \rightarrow \infty$.
Maximum mixing occurs at higher momenta as more Landau levels are filled. We caution, however, that mixing can 
be much larger for even modest increases in $\beta$.  For example, we find mixing of up to 70\% at intermediate
momenta for $\beta = 2.2$.

  The peaks and valleys apparent in the exciton dispersions of Figures \ref{fig:dispersion}
and \ref{fig:dispersion_gne4} are of
particular interest, because they lead to van Hove singularities in the collective
mode density of states.  Such singularities are in principle detectable in
inelastic light scattering \cite{pinczuk} and, via coupling through disorder, microwave
absorption \cite{KallandHalp2}.  Thus it is useful to know when and
where such singularites might be found.  To compute them more accurately,
we now turn to the many-body formulation of this problem.

\section{Many-body problem}
\label{sec:many-body}

We study the many-body exciton dispersion by diagonalizing the many-body
interacting Hamiltonian on a subspace of trial wavefunctions, which consist 
of a single electron excited from the Fermi sea.  In the 2DEG, this 
procedure is known to give results for the exciton dispersion 
that coincide with those obtained from the 
poles of response functions computed in a generalized random phase approximation (RPA)
\cite{RPA}. Indeed, these wavefunctions are exact in the non-interacting case, 
and should thus provide a good basis for the weakly interacting regime.  On the other
hand, the validity of these wavefunctions may persist for stronger interactions if the 
Landau level mixing is weak.  Experience has shown that in the 2DEG, 
the degree of Landau level mixing is considerably less than one would naively anticipate
based on the ratio of potential and kinetic energies.  This mixing is known 
to be small \cite{MilalekFertig} even when this ratio is of order unity.

In the case of graphene, this notion is supported by the results of the previous section.
In the two-body calculation, where we take $\beta = 0.73$, we find that in the lowest exciton state, 
Landau level mixing occurs at a low level,
causing the inflections in the dispersion to become somewhat more pronounced, but otherwise
having little qualitative effect.  
From this we conclude that reasonably accurate results for the lowest exciton states may be obtained using 
only those electron and hole Landau levels nearest the Fermi energy.  Thus, we focus our attention
on exciton states in which the Landau level indices of the electron and hole differ by 
at most $l_z=1$.

As in the previous section, we make the approximation that the Coulomb interaction has SU(2) symmetry
with respect to the two sublattices. 
In this approximation, the Coulomb interaction conserves pseudospin in addition to the usual 
spin conservation.  As a result, exciton states mixed by the Coulomb interaction 
have the same net $s_z$ and $t_z$ and consequently have the same net contribution to the non-interacting
energy from Zeeman and pseudospin splitting.

We restrict our consideration of mixing to those exciton states with 
the same non-interacting energy.  In fact, such degeneracy is almost always due to 
the several possible spin and 
pseudospin configurations of the individual electron and hole when the net $s_z$ or $t_z$ is zero.
Due to the nonuniform spacing of Landau levels in graphene, 
two electron-hole pairs with different Landau level indices will not have the same kinetic energy 
unless one of the pairs has $l_z > 1$. The only exception occurs when $\gamma_0 < 4$, 
in which case the non-interacting states $\ket{1,0}$ and $\ket{0,-1}$ have the same kinetic energy.  
Under these considerations, we concentrate on the cases $\gamma_0 < 4$, $\gamma_0 = 4$, 
$\gamma_1 = 4$, and $\gamma_2 = 4$. In particular, the first case offers the possibility of Landau
level mixing for $l_z = 1$ and intra-level spin and pseudospin waves for $l_z = 0$.

At this level of approximation, the dispersion relation $E(P)$ for an exciton can be written
\begin{eqnarray}
E(P) = E_{\rm kin} - |g \mu_B B| s_z - \hbar \omega_t\: t_z + \Delta E^{\nu}(P),
\end{eqnarray}
where $E_{\rm kin}$, $s_z$, and $t_z$ are, respectively, the combined kinetic energy, spin, and pseudospin of the 
electron and hole.  The superscript $\nu$ provides an arbitrarily assigned index for the 
several branches of the additional contribution $\Delta E^{\nu} (P)$ due to interactions.

In our many-body approach, $\Delta E$ has three contributions.
The first is the direct Coulomb interaction between the excited electron and hole, whose matrix elements
are identical to those found in the two-body approach of the previous section.  This ``excitonic effect'' 
can be obtained in a diagrammatic expansion by including vertex corrections \cite{KallHalp}. 
In our continuum approximation, 
this contribution only scatters between exciton states if the electron and hole 
individually have the same $s_z$ and $t_z$ in the initial and final state. 
The second is the exchange term or ``depolarization effect,'' which is the only contribution 
present in RPA.
Physically, it represents the annihilation of 
the electron-hole pair at one point in space and its recreation at another point.
This process only affects excitons with vanishing net $s_z$ and $t_z$.
Finally, each single-particle sublevel has an associated 
self-energy which arises from exchange interactions with those occupied sublevels in the Fermi sea
having the same spin and pseudospin.
This effect provides a $P$-independent shift to the exciton dispersion, guaranteeing
that its energy is always non-negative.  It also pushes the $P \rightarrow \infty$ 
limit above the non-interacting energy.  

\begin{figure}
\includegraphics[width=\columnwidth]{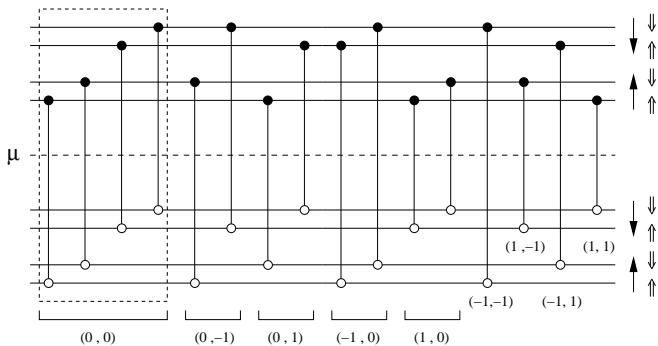}
\caption{Exciton states for the case $\gamma = 4$, with definite spin
and pseudospin denoted $(s_z,t_z)$.  The dashed box indicates states 
admixed by exchange interactions.  Direct Coulomb interactions do not admix these states.
}
\label{fig:g4}
\end{figure}

The situation is illustrated in Figure \ref{fig:g4} for the simplest
case $\gamma_n=4$. There are 16 exciton states with minimal kinetic energy which are  
differentiated by the spin and pseudospin sublevels of the electron and hole.  
Direct interactions only 
scatter electrons (holes) into electron (hole) sublevels with the same $s_z$ and $t_z$,
and thus do not cause mixing.  Exchange interactions scatter among the 4 states with $s_z = t_z = 0$.

\begin{figure}
\includegraphics[width=\columnwidth]{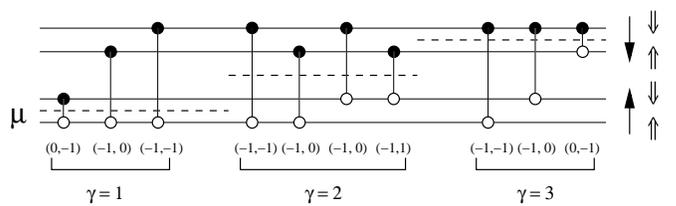}
\caption{Exciton states for the case $l_z = 0$, $\gamma < 4$, labeled by spin and pseudospin 
as $(s_z,t_z)$. 
}
\label{fig:LL_lz0}
\end{figure}

The case $\gamma < 4$, $l_z = 0$, is shown in Figure \ref{fig:LL_lz0}. The lowest exciton states
actually have zero kinetic energy; the excitation is a pure spin or pseudospin wave within
the Landau level.  Neither the direct nor exchange interactions mix the states. 
The dispersions are identical to that of the two-body case up to a constant due to 
the exchange self-energy correction as well as Zeeman and pseudospin splitting.

\begin{figure}
\includegraphics[width=\columnwidth]{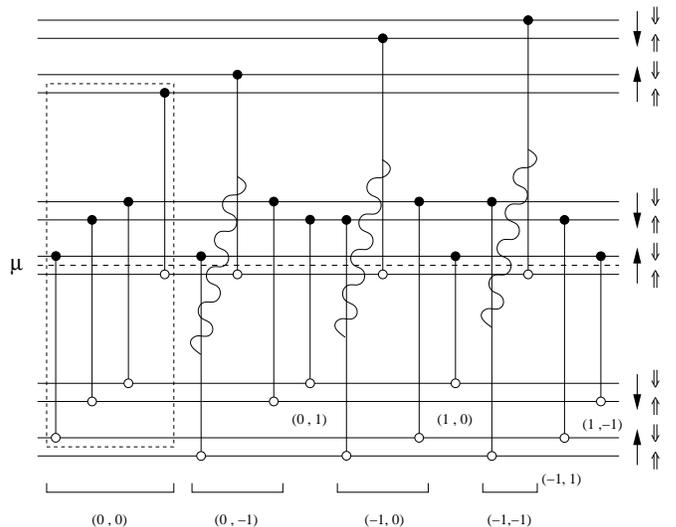}
\caption{Exciton states for the case $\gamma_0 = 1$, $l_z = 1$, with definite spin
and pseudospin denoted $(s_z,t_z)$.  The dashed box indicates states 
admixed by exchange interactions, wavy lines show states coupled by the direct Coulomb interaction.
}
\label{fig:g1}
\end{figure}

\begin{figure}
\includegraphics[width=\columnwidth]{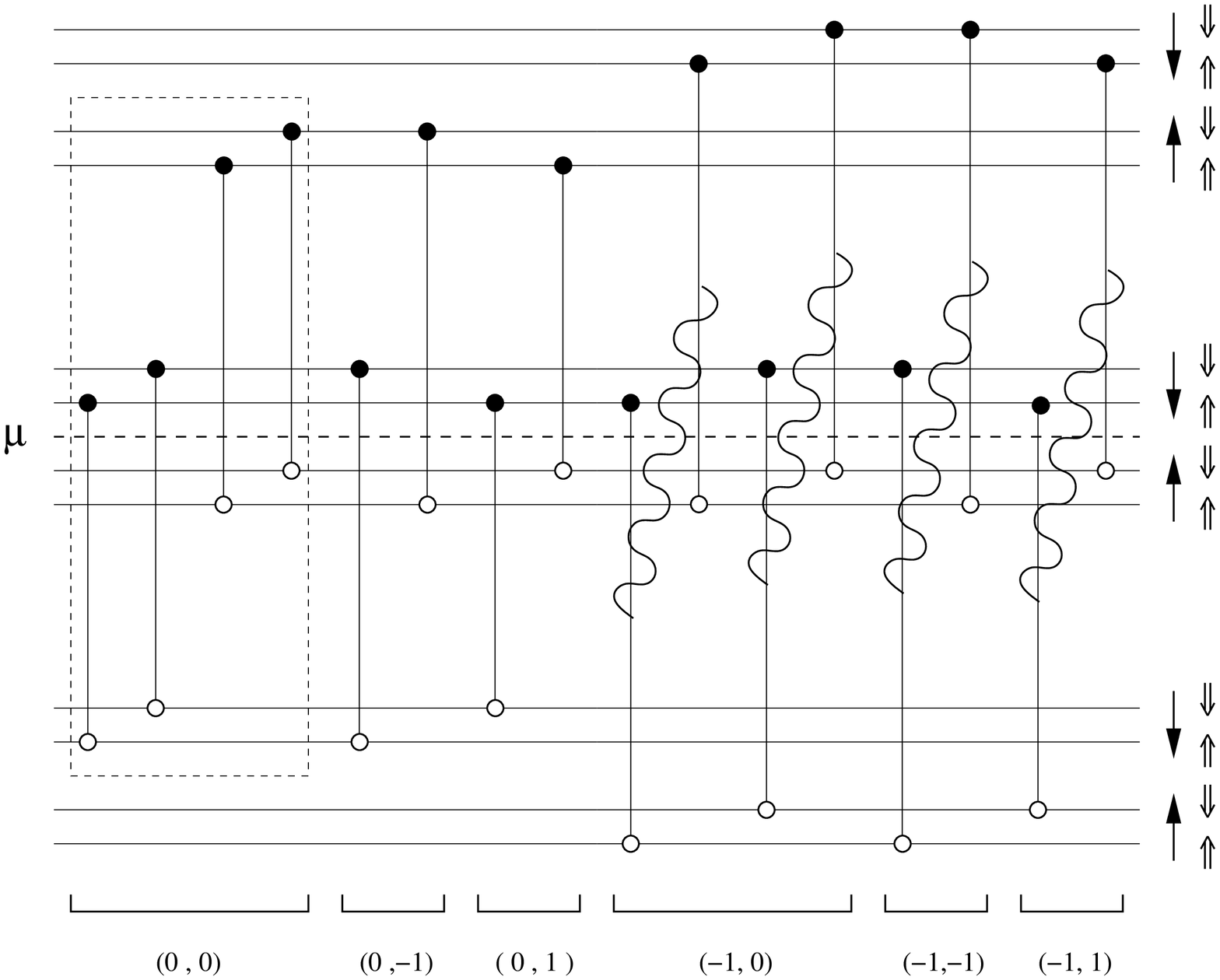}
\caption{Exciton states for the case $\gamma_0 = 2$, $l_z = 1$, with definite spin
and pseudospin denoted $(s_z,t_z)$.  The dashed box indicates states 
admixed by exchange interactions, wavy lines show states coupled by the direct Coulomb interaction.
}
\label{fig:g2}
\end{figure}

\begin{figure}
\includegraphics[width=\columnwidth]{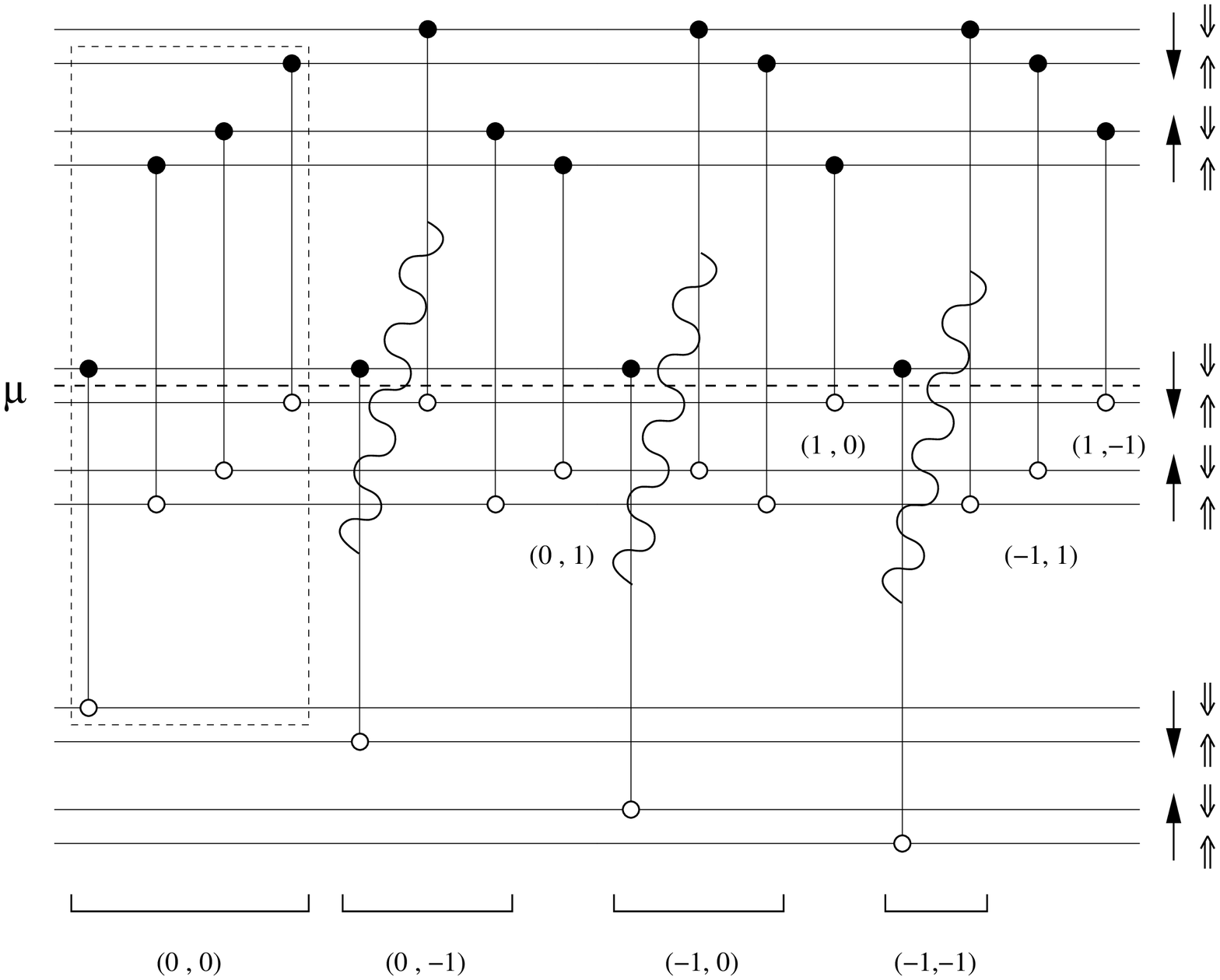}
\caption{Exciton states for the case $\gamma_0 = 3$, $l_z = 1$, with definite spin
and pseudospin denoted $(s_z,t_z)$.  The dashed box indicates states 
admixed by exchange interactions, wavy lines show states coupled by the direct Coulomb interaction.
}
\label{fig:g3}
\end{figure}

The most interesting scenario is illustrated in Figures \ref{fig:g1}-\ref{fig:g3} for the 
cases $\gamma_0 = 1,2,3$. 
Since the levels $n=-1,0,1$ are equally spaced, there 
are excitons with $l_z = 1$ whose kinetic energies are degenerate and may be split
by Landau level mixing.  This is
caused by the direct Coulomb interaction, which admixes states connected by a wavy line. 
As in the $\gamma_n = 4$ case, the $s_z = t_z = 0$ excitons are mixed by the exchange interaction.
Note that the number of exciton states with given values of $s_z$ and $t_z$
is closely tied to the number of filled sublevels. 

We now derive the three many-body contributions by diagonalizing the 
Hamilonian on the subspace of trial wavefunctions. 
The matrix elements between the non-interacting exciton states are 
\begin{align}
M_{1,2;1',2'} &\equiv 
\bra{\Omega}(a_{1'}^{\dagger} a_{2'})^{\dagger} \hat{H}
a^{\dagger}_1 a_2 \ket{\Omega} 
\\ \nonumber & \quad 
- \bra{\Omega} \hat{H} \ket{\Omega} \delta_{1,1'} \delta_{2,2'}, 
\end{align} 
where $\ket{\Omega}$ is the filled Fermi sea.
Diagonalizing $\hat{H}$ on the subspace of trial wavefunctions
may be viewed as a two-body problem with Hamiltonian $\hat{M}$.
The many-body Hamiltonian has the form
\begin{eqnarray}
H &=& \mathop{\sum_1} \varepsilon_1 a^{\dagger}_1 a_1
+ \frac{1}{2} \mathop{\sum_{1,2,3,4}} V_{1,2,3,4}
a_1^{\dagger}
a_2^{\dagger}
a_3
a_4 ,
\end{eqnarray}
where $V_{1,2,3,4} = V_{2,1,4,3}$ due to the inversion symmetry of the Coulomb interaction.
This leads to $\hat{M} = \hat{M}^0 + \hat{M}^{d} + \hat{M}^{x} + \hat{M}^{\Omega}$ 
where 
\begin{eqnarray}
M^0_{1,2;1',2'} &=& \delta_{1,1'} \delta_{2,2'} (\varepsilon_1 - \varepsilon_2) \\
M^{d}_{1,2;1',2'} &=& - V_{1',2,2',1} \\
M^{x}_{1,2;1',2'} &=&  V_{1',2,1,2'} \\
M^{\Omega}_{1,2;1',2'} &=& \mathop{\sum_3} \Theta(\mu - \varepsilon_3)
[ \delta_{1,1'} V_{3,2,3,2'} 
\\ & & \nonumber \quad - \delta_{2,2'} V_{1',3,1,3}] . 
\end{eqnarray}
These terms represent the noninteracting, direct, exchange, and exchange self-energy contributions, 
respectively.
We assume that the direct interaction with the Fermi sea is cancelled by a neutralizing background charge.
The calculation of these matrix elements is greatly simplified by separating the relative and center-of-mass
motion of the electron and hole by means of the canonical transformation developed for the two-body 
problem. The computation of matrix elements and $\Delta E (P) $ 
is discussed extensively in the Appendix.

For a given chemical potential, interactions give rise to several branches 
$\Delta E^{\nu}(P)$. Since there are fewer distinct branches 
than there are states (16 states in the case $l_z = 1$, 4 for $l_z = 0$,)
some of the exciton states remain degenerate.  Some of this degeneracy may remain even after 
pseudospin and Zeeman splitting are taken into account, which is readily done by inspection 
of the $s_z$ and $t_z$ quantum numbers in Table \ref{tab:qnumbers}.  Based on an analysis of Figures
\ref{fig:g4}-\ref{fig:g3} and of the matrix elements, we can deduce the different spin 
and pseudospin quantum numbers carried by excitations with the same $\Delta E (P)$, 
as is shown in Table \ref{tab:qnumbers}.  Analytic expressions for these branches are
given in the Appendix in Table \ref{tab:dispersions}.

Along with $s_z$ and $t_z$, we may also consider 
total spin $\hat{\bm{s}}^2 = S(S+1)$
and total pseudospin $\hat{\bm{t}}^2 = T(T+1)$.
The Fermi sea must be in a spin (pseudospin) singlet state 
in order for $S$ ($T$) to be a meaningful quantum number for the excitons,
as has been pointed out previously for the 2DEG \cite{KallHalp}.
Consequently, excitons have well-defined $S$ only when $\gamma_n = 4$ and 
well-defined $T$ only when $\gamma_n = 2,4$.  

From symmetry considerations, we can predict an effect of intervalley scattering.
Such scattering will not pick out a preferred direction in pseudospin space, but may
lift any degeneracy between pseudospin singlet and triplet states.  Based on Table 
\ref{tab:qnumbers}, this effect should be easiest to discern in the case 
$\gamma_0 = 2$.  In this case, there are spin-carrying excitations 
(both for the intra-level exciton as well as branches $\Delta E^3$ and 
$\Delta E^4$ of the $l_z = 1$ exciton) having four-fold degeneracy purely due to pseudospin
(provided $\hbar \omega_t = 0$.)
An estimate on the strength of intervalley scattering could be gleaned from 
any splitting observed between the $T=0$ and $T=1$ states.

\begin{table}[htp]
\begin{center}
\begin{tabular}{|c|l|r|r|r|r|}
\hline
filling & branch & $s_z$ & $S$ & $t_z$ & $T$  \\
\hline
$\gamma_0 = 2$, $\:l_z = 0$ & $\: \Delta E^0$ & -1 & - & all & all \\
\hline
$\gamma_0 = 1,3$; $\:l_z = 0$ & $\: \Delta E^0$ 
 & 0 & - & -1 & - \\
 & & -1 & - & 0 & - \\
 & & -1 & - & -1 & - \\
\hline
$\gamma_n = 4$, $\:l_z = 1$ & $\: \Delta E^1$ & 0 & 1 & 0 & 1 \\
 & $\: \Delta E^2$ & all & all & all & all \\
\hline
\end{tabular}
\end{center}
$\begin{array}{lcr}
\begin{minipage}{3 cm}
\centerline{$\gamma_0 = 2, l_z = 1$}
\vskip 0.2 cm
\begin{tabular}{|l|r|r|r|r|}
\hline
branch & $s_z$ & $S$ & $t_z$ & $T$  \\
\hline
$\: \Delta E^1$ & 0 & - & 0 & 1  \\
$\: \Delta E^2$ & 0 & - & 0 & 1 \\
 & 0 & - & 0 & 0 \\
 & 0 & - & 0 & 0 \\
 & 0 & - & -1 & 1 \\
 & 0 & - & -1 & 1 \\
 & 0 & - & +1 & 1 \\
 & 0 & - & +1 & 1 \\
 $\: \Delta E^3$ 
   & -1 & - & 0 & 0 \\
 & -1 & - & 0 & 1 \\
 & -1 & - & -1 & 1 \\
 & -1 & - & +1 & 1 \\
 $\: \Delta E^4$ 
   & -1 & - & 0 & 0 \\
 & -1 & - & 0 & 1 \\
 & -1 & - & -1 & 1 \\
 & -1 & - & +1 & 1 \\
\hline
\end{tabular}
\end{minipage}
& \hskip 0.5 cm &
\begin{minipage}{3 cm}
\centerline{$\gamma_0 = 1,3; l_z = 1$
}
\vskip 0.2 cm
\begin{tabular}{|l|r|r|r|r|}
\hline
branch & $s_z$ & $S$ & $t_z$ & $T$  \\
\hline
$\: \Delta E^1$ & 0 & - & 0 & - \\
$\: \Delta E^2$ & 0 & - & 0 & - \\
& 0 & - & 0 & - \\
& 0 & - & 0 & - \\
& 0 & - & -1 & - \\
& 0 & - & +1 & - \\
& -1 & - & 0 & - \\
& -1 & - & +1 & - \\
& +1 & - & 0 & - \\
& +1 & - & -1 & - \\
$\: \Delta E^3$ 
  & 0 & - & -1 & - \\
& -1 & - & 0 & - \\
&  -1 & - & -1 & - \\
$\: \Delta E^4$ 
  & 0 & - & -1 & - \\
& -1 & - & 0 & - \\
&  -1 & - & -1 & - \\
\hline
\end{tabular}
\end{minipage}
\end{array}
$
\caption{Quantum numbers of the various exciton branches
for different numbers of filled sublevels. Dash indicates
the quantum number is not well-defined. For $\gamma_n = 4$, branch
$\Delta E^2$ contains all the spin and pseudospin states not contained in branch $\Delta E^1$.
In the $\gamma_0 = 2$, $l_z = 0$ case, all 4 pseudospin states are present.
}
\label{tab:qnumbers}
\end{table}


Figures \ref{fig:lz0} - \ref{fig:gamma0=1234} plot the various branches $\Delta E^{\nu} (P)$
for the different cases. The $l_z=0$ case (intra-level exciton, Figure
\ref{fig:lz0}) is the simplest; 
it has a single branch $\Delta E^0$ given by the two-body result
plus a constant exchange self-energy correction.  This correction renders the mode
gapless (if we ignore Zeeman and 
pseudospin contributions to the non-interacting energy,) and we recognize it as the quadratic 
Goldstone mode corresponding to broken spin and pseudospin symmetries. 
Next in complexity are the cases $\gamma_1 = 4$ and $\gamma_2 = 4$, 
which include an exchange interaction among the $s_z = t_z = 0$ states.
This interaction splits off the totally symmetric combination with 
$s_z = t_z = 0$ and $S = T = 1$ in branch $\Delta E^1$,
and the other 15 states are in branch $\Delta E^2$.  The former
has a nonmonotonic character (particularly for larger $n$,) 
while the latter is essentially monotonic.

Figure \ref{fig:gamma0=1234} shows the most interesting case $n=0$.
The case $\gamma_0 = 4$ is similar in all respects to the other 
cases $\gamma_n = 4$, having 15 states in branch $\Delta E^2$ and a 
single state pushed up by exchange interactions 
into branch $\Delta E^1$.  When less than four sublevels are filled, two 
additional branches appear due to the possibility of Landau level mixing.
In general, we expect the cases $\gamma_n = 1,2,3$ to have different $P$ dependences 
for $\Delta E^{\nu} (P)$ since the exchange self-energy parameters 
(which generally do not simply add a constant shift to the energy) depend on the number of filled sublevels.
However, for $n = 0$, there is a degeneracy between exchange constants 
which is protected by particle-hole symmetry (this point is elaborated in the Appendix,)
and the branches $\Delta E^{\nu}(P)$ are identical for these three cases.  
This feature is peculiar to graphene has no analog in the 2DEG. 

On the other hand, 
the number and type of excitations in each branch does depend on the particular
value of $\gamma_0$, as can be seen in Table \ref{tab:qnumbers}.
Note that if we were to add the constant Zeeman contribution to Figures \ref{fig:lz0}-\ref{fig:gamma0=1234}, 
we would see branch $\Delta E^2$ separate into three evenly spaced levels for the cases 
$\gamma_0 = 1,3$ and $\gamma_n = 4$, 
but not $\gamma_0 = 2$. For $\gamma_0 = 1,3$ we would also see branches $\Delta E^0$, $\Delta E^3$, 
and $\Delta E^4$ split into two levels.
These effects can be deduced from the $s_z$ values listed in Table \ref{tab:qnumbers}.
Similar considerations apply to the much smaller pseudospin splitting. 
Note that even after the inclusion of these effects, the dispersions of the $\gamma_0=1,3$ cases 
are identical, due physically to particle-hole symmetry.

We remark that the only $P$-dependent contribution to the dispersion
not captured in the two-body analysis appears
in branch $\Delta E^1$, present in the $l_z = 1$ excitons. 
Here, exchange interactions lead to dramatic differences, for example
as is evident in Figure \ref{fig:gamma2=4}.  The very high peaks apparent there not only do not
appear in the two-body formulation, they are quantitatively much larger than
analogous many-body corrections for the 2DEG \cite{KallHalp}.  This
relatively large effect may be traced to the unusually large number of pair
excitations coupled together in the graphene system.  It would be most
interesting if the peak/dip structure evident in Figures \ref{fig:gamma1=4} and 
\ref{fig:gamma2=4} could be observed in inelastic light scattering or optical 
absorption.\cite{Sadowski}

\begin{psfrags}
\psfrag{xaxis}{$P \ell$}
\psfrag{yaxis}[Bc][Bc]{$\Delta E \: (e^2 / \epsilon \ell)$}
\psfrag{dE0}{$\Delta E^0$}
\psfrag{dE1}{$\Delta E^1$}
\psfrag{dE2}{$\Delta E^2$}
\psfrag{dE3}{$\Delta E^3$}
\psfrag{dE4}{$\Delta E^4$}
\begin{figure}
\center{
\includegraphics[width=\columnwidth]{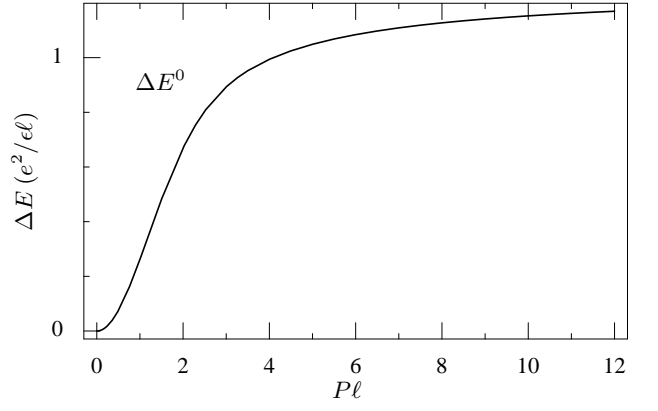}
}
\caption{Energy shift $\Delta E^1$ for the intra- Landau level excitons in the case $\gamma_0 < 4$.
}
\label{fig:lz0}
\end{figure}

\begin{figure}
\center{
\includegraphics[width=\columnwidth]{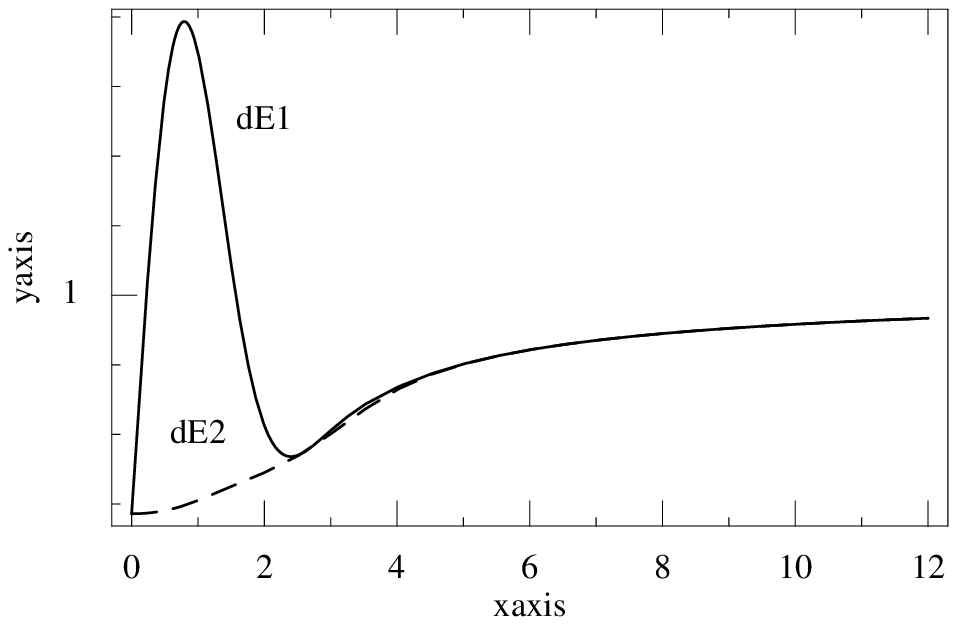}
}\caption{Energy shifts $\Delta E^1$ (solid) and $\Delta E^2$ (dashed) for $\gamma_1 = 4$.}
\label{fig:gamma1=4}
\end{figure}

\begin{figure}
\center{
\includegraphics[width=\columnwidth]{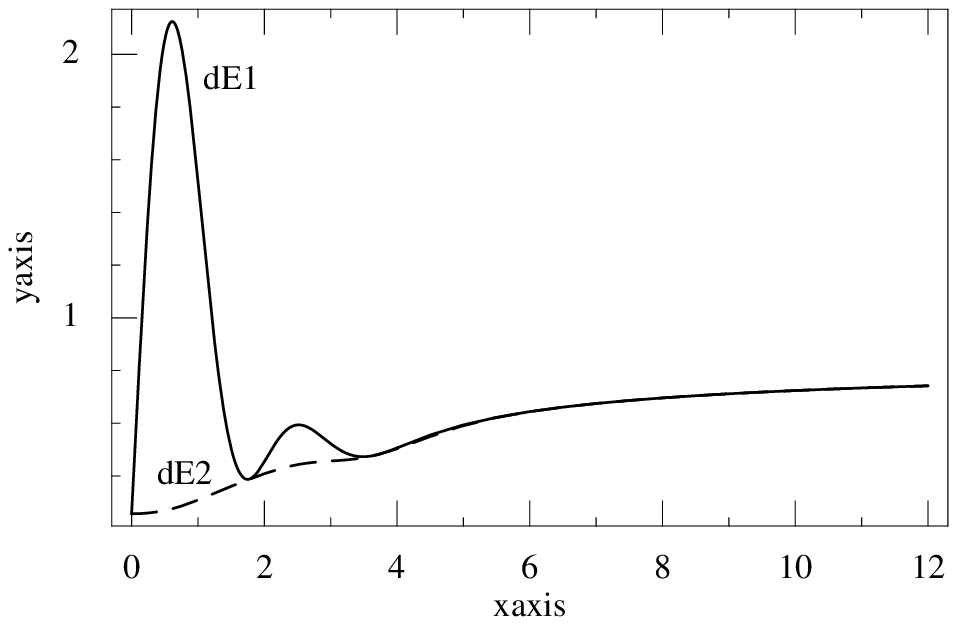}
}\caption{Energy shifts $\Delta E^1$ (solid) and $\Delta E^2$ (dashed) for $\gamma_2 = 4$.}
\label{fig:gamma2=4}
\end{figure}

\begin{figure}
\center{
\includegraphics[width=\columnwidth]{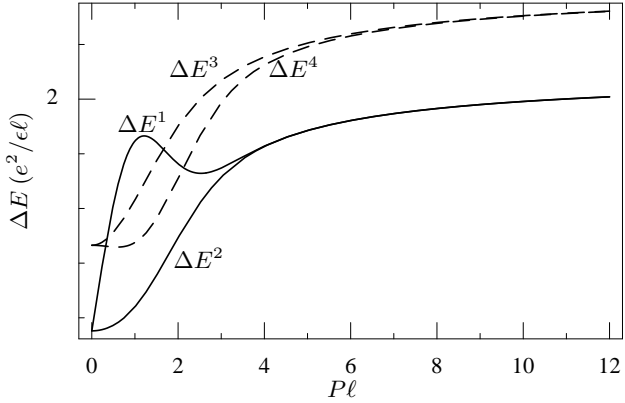}
}
\caption{Energy shifts for inter- Landau level excitons in the case $n=0$.
When $\gamma_0 = 4$, only $\Delta E^1$ and $\Delta E^2$ (solid) are present.
For $\gamma_0 < 4$, the additional branches $\Delta E^3$ and $\Delta E^4$ (dashed) appear.
}
\label{fig:gamma0=1234}
\end{figure}
\end{psfrags}

\section{Conclusions}
In this article, we have analyzed the excitations from filled Landau levels in 
graphene using two approaches.  In the first, we tackle the two-body 
problem for the excited particle and hole in a magnetic field.  
We introduce a canonical transformation which separates the relative motion and 
allows us to identify the symmetries and conservation laws for the system.
We then compute the dispersion of excitons as a function of $P$, restricting the
available electron and hole levels based on the value of the chemical potential.  
By including higher Landau levels in the calculation, we establish that
Landau level mixing is weak for the lowest exciton states even though the estimated
ratio $\beta$ of potential and kinetic energies is of order 1.  We caution, however, 
that mixing can become important for even modest increases above the value $\beta = 0.73$
used here.

Armed with this insight, we turn to the many-body problem, where we consider the implications of 
the spin and pseudospin sublevels.
We find that the large number of degenerate states coupled together by exchange interactions 
gives rise to an excitation with enhanced many-body corrections and strong peak/dip
features in its dispersion, which could be observed experimentally. A careful analysis of the spin and valley quantum 
numbers shows the sensitivity of the spectrum to the number of filled sublevels $\gamma_0$ in Landau level
$n = 0$.  In particular, when $\gamma_0 < 4$, 
there are new branches in the dispersion relation (due both to Landau level mixing
and to intra-level excitations) 
which do not appear when $\gamma_0 = 4$.  We also find that the case $\gamma_0 = 2$ 
is distinguished from the other cases by the absence of any Zeeman splitting of the low-lying
exciton states.
The cases $\gamma_0 = 1,3$ are identical due to particle-hole symmetry effects which have no
analog in the 2DEG.

\acknowledgments
The authors would like to thank A. Castro Neto for stimulating discussions on graphene.
This work was supported through the NSF by Grant No. DMR-0454699, and by MAT2005-07369-C03-03 (Spain).

\begin{appendix}
\section{}
The oscillator wavefunctions for the one-particle basis defined in the text are
\begin{align}
\varphi_n(x) & \equiv (\sqrt{\pi} \: 2^n n!)^{-1/2}\: H_n(x)\: e^{-x^2/2} , 
\end{align}
where $H_n$ are Hermite polynomials. 
The two-dimensional oscillator wavefunctions are obtained
by the repeated application of the raising operators $c_{\pm}^{\dagger}$ 
to the ground state. Writing $\Phi_{0,0}(z,\bar{z}) = (2 \pi)^{-1/2} e^{- z \bar{z} / 4}$
with $z = x + i y$,  
we have $\Phi_{n_1,n_2} \propto [c_+^{\dagger}]^{n_1} [c_-^{\dagger}]^{n_2} \Phi_{0,0}$, or 
explicitly
\begin{align}
\Phi_{n_1,n_2} (\bm{r}) &\equiv
 (2 \pi)^{-1/2} 
(2^{n_1 + n_2} n_1 ! \: n_2 ! \: )^{-1/2} \\ \nonumber 
& \quad \quad (- 2 \partial_z + \bar{z} / 2)^{n_1}
(2 \partial_{\bar{z}} - z / 2)^{n_2}
\: e^{- z \bar{z} / 4}
\\ \nonumber &=
(2 \pi )^{-1/2} 2^{-|l_z|/2} \frac{n_- !}{\sqrt{n_1 ! n_2 !}}
\: e^{- i \: l_z \phi} \sgn(l_z)^{l_z}
\\ & \quad \quad r^{|l_z|} 
L_{n_-}^{|l_z|}(r^2 / 2) e^{- r^2 / 4} ,
\nonumber 
\end{align}
where $L$ denotes Laguerre polynomials, $z = x + i y$, $l_z = n_1 - n_2$, $n_- = \min(n_1,n_2)$, $z / |z| = e^{i \phi}$
and $\sgn(l_z)^{l_z} \rightarrow 1$ for $l_z = 0$.
The one- and two-dimensional oscillator wavefunctions are connected by the relation
\begin{align}
\Phi_{n_1,n_2}(\bm{r}) &= (-1)^{n_1 + n_2} 
\int \frac{dX}{\sqrt{2\pi}} e^{i X y} 
\label{eq:harmonic}
\\ \nonumber & \quad\quad \varphi_{n_1} (X - x/2) \varphi_{n_2} (X + x/2) .
\end{align}

The separation of the relative coordinate of the two-body state can be achieved 
with an appropriate choice of the two-body basis.  We choose the basis in which the electron and 
hole lie in Landau levels $n_1$ and $n_2$, respectively, and the pair has definite momentum $\bm{P}$.
For the 2DEG, the one-body eigenstates are defined
\begin{align}
\psi^{\rm 2DEG}_{n,k} (\bm{r}) &\equiv L_y^{-1/2} e^{i k y} \varphi_n(x - k) .
\end{align}
The two-body wavefunctions are formed using a special linear combination 
of one-body states for the electron and hole:
\begin{align}
\Psi_{\bm{P}; n_1, n_2}^{\rm 2DEG} (\bm{r}_1 , \bm{r}_2) &\equiv L_y \int \frac{dk}{\sqrt{2 \pi}}
\:e^{i P_x k} \\ \nonumber
& \quad \quad \psi^{*\: \rm 2DEG}_{n_2, - \frac{1}{2} P_y + k}( \bm{r}_2 )
\psi^{\rm 2DEG}_{n_1, \frac{1}{2} P_y + k}( \bm{r}_1 ) \\
&= \nonumber e^{i X y} e^{i \bm{P} \cdot \bm{R}}
\Phi_{n_1, n_2}(\bm{r} + \hat{z} \times \bm{P}) , 
\end{align}
where $\bm{r}$ and $\bm{R}$ are the relative and center-of-mass coordinates corresponding to 
$\bm{r}_1$ and $\bm{r}_2$.  
This basis implements the canonical transformation developed for the two-body problem.
The second equality can be established using Eq. \ref{eq:harmonic}.

The same transformation connects the one- and two-body 
bases for graphene, which we review here in component form.
For the one-body states,  
\begin{eqnarray}
[\psi_{n,k}^t(x,y)]_{\tau} &\equiv& L_y^{-1/2} e^{i k y} 
[\sqrt{2}]^{\delta_{n,0}-1} 
\\ & & \nonumber 
s(n,t,\tau) \varphi_{\lambda(n,t,\tau)}(x - k),
\end{eqnarray}
and for the two-body states,
\begin{eqnarray}
& &[\Psi_{\bm{P}; n_1, n_2; t_1, t_2} (\bm{r}_1 , \bm{r}_2)]_{\tau_1 , \tau_2} 
\equiv L_y \int \frac{dk}{\sqrt{2 \pi}}
\:e^{i P_x k} \\ \nonumber
& & \quad \quad \quad [\psi^{t_2}_{n_2, - \frac{1}{2} P_y + k}( \bm{r}_2 )]_{\tau_2}^*
[\psi^{t_1}_{n_1, \frac{1}{2} P_y + k}( \bm{r}_1 )]_{\tau_1} 
\\ \nonumber
&=& \quad e^{i X y} e^{i \bm{P} \cdot \bm{R}} [\sqrt{2}]^{\delta_{n_1,0} + \delta_{n_2,0} - 2}
s(n_1,t_1,\tau_1)  s(n_2,t_2,\tau_2) 
\label{eq:CMform} \\ 
& & \quad \cdot \: \Phi_{\lambda(n_1,t_1,\tau_1),\lambda(n_2,t_2,\tau_2)}(\bm{r} + \hat{z}\times \bm{P}),
\end{eqnarray}
where $t_1$ and $t_2$ denote the pseudospin index of the electron and hole sublevels.

The separation of the center-of-mass and relative coordinates in the latter expression is extremely useful 
in the calculation of matrix elements of $\hat{M}$.  
The matrix elements $M_{1,2;1',2'}$ represent the scattering between initial and final 
states for the electron-hole pair.
Let $n_1$, $n_2$, $n_1'$, $n_2'$ denote the Landau level indices of the 
initial electron, initial hole, final electron, and final hole states, respectively. 
Similar conventions denote the initial and final momentum $\bm{P}$ of pair as well as the 
spin $s$ and pseudospin $t$ of the individual electron and hole.
The matrix elements are
\begin{eqnarray*}
M^{d}_{1,2;1',2'} &=& \nonumber
\delta_{s_1,s_1'} \delta_{s_2,s_2'} \delta_{t_1,t_1'} \delta_{t_2,t_2'}
\mathop{\sum_{\tau_1,\tau_2}} \int \: d\bm{r}_1 \: d\bm{r}_2 
\\ & & \nonumber u(\bm{r}_1 - \bm{r}_2)
[\Psi_{\bm{P}'; n_1', n_2' ; t_1' t_2'} (\bm{r}_1,\bm{r}_2)]_{\tau_1,\tau_2}^*
\\ & & \quad [\Psi_{\bm{P}; n_1, n_2 ; t_1 t_2} (\bm{r}_1,\bm{r}_2)]_{\tau_1,\tau_2} \\
M^{x}_{1,2;1',2'} &=& \nonumber
\delta_{s_1,s_2} \delta_{s_1',s_2'} \delta_{t_1,t_2} \delta_{t_1',t_2'}
\mathop{\sum_{\tau_1,\tau_2}} \int \: d\bm{r}_1 \: d\bm{r}_2 
\\ & & \nonumber u(\bm{r}_1 - \bm{r}_2)
[\Psi_{\bm{P}'; n_1', n_2' ; t_1' t_2'} (\bm{r}_1,\bm{r}_1)]_{\tau_1,\tau_1}^*
\\ & & \quad
[\Psi_{\bm{P}; n_1, n_2 ; t_1 t_2} (\bm{r}_2,\bm{r}_2)]_{\tau_2,\tau_2} \\
M^{\Omega}_{1,2;1',2'} &=& \delta_{1,1'} m^{\Omega}_{2,2'} - \delta_{2,2'} m^{\Omega}_{1,1'}, \\
m^{\Omega}_{1,1'} &\equiv&  \nonumber
\mathop{\sum_{n}} \Theta [\mu - \varepsilon(n, s_1, t_1)] 
\:\delta_{s_1, s_1'} \delta_{t_1, t_1'} 
\\ & & \nonumber 
\int \frac{d k}{2 \pi} \: 
\mathop{\sum_{\tau_1, \tau_2}} \int \: d\bm{r}_1 \: d\bm{r}_2 \: u(\bm{r}_1 - \bm{r}_2) 
\\ & & 
[\psi_{n_1',k_1'}^{t_1}(\bm{r}_1)]_{\tau_1}^*
[\psi_{n,k}^{t_1}(\bm{r}_2)]_{\tau_2}^*
\\ & & \nonumber \quad \quad
[\psi_{n_1,k_1}^{t_1}(\bm{r}_2)]_{\tau_2}
[\psi_{n,k}^{t_1}(\bm{r}_1)]_{\tau_1},
\end{eqnarray*}
where $u(\bm{r}) = \beta / r$ is the Coulomb potential.
The integrals are straightforward after changing to relative and center-of-mass coordinates.
For the direct ($M^{d}$) and exchange ($M^{x}$) contributions, 
the use of Eq. \ref{eq:CMform} leads immediately to expressions involving $\Phi$, whereas
for the Fermi sea term ($M^{\Omega}$) this requires the use of Eq. \ref{eq:harmonic}.

It is useful to define 
\begin{align}
u_{n_1 , n_2 ; n_1', n_2'} (\bm{P}) &\equiv
\int d\bm{r}  \:u(\bm{r} - \hat{z} \times \bm{P})
\Phi_{n_1',n_2'}^*(\bm{r}) \Phi_{n_1,n_2}(\bm{r}) 
\\ v_{n_1 , n_2 ; n_1', n_2'} (\bm{P}) &\equiv \hat{u}(\bm{P}) 
\Phi_{n_1',n_2'}^*(\hat{z} \times \bm{P}) \Phi_{n_1,n_2}(\hat{z} \times \bm{P}) ,
\end{align}
where $\hat{u}(\bm{k}) = 2\pi \beta / k$ is the Fourier transform of the Coulomb potential.
The quantities $-u$ and $v$ are, in fact, the direct and exchange matrix elements for the 2DEG.
The graphene matrix elements are then
\begin{eqnarray*}
M^0_{1,2;1',2'} &=& \delta_{1,1'} \delta_{2,2'} (\varepsilon_1 - \varepsilon_2) \\
M^{d}_{1,2;1',2'} &=& 
\delta_{\bm{P},\bm{P}'} 
\delta_{s_1, s_1'} \delta_{t_1,t_1'} \delta_{s_2,s_2'} \delta_{t_2,t_2'}
m^{d}_{1,2;1',2'} (\bm{P}) \\
m^{d}_{1,2;1',2'} (\bm{P}) &\equiv& 
- \frac{1}{4} [\sqrt{2}]^{\delta_{n_1,0} + \delta_{n_1',0} + \delta_{n_2,0} + \delta_{n_2',0}}
\\ & & \nonumber 
\mathop{\sum_{\mu,\nu = 0}^1} 
s_{\mu}(n_1) s_{\nu}(n_2) s_{\mu}(n_1') s_{\nu}(n_2')
\\ & & \nonumber 
\quad u_{
\lambda_{\mu}(n_1), \lambda_{\nu}(n_2); \lambda_{\mu}(n_1'), \lambda_{\nu}(n_2')
}(\bm{P})
\\ 
M^{x}_{1,2; 1', 2'} &=&
\delta_{\bm{P},\bm{P}'} 
\delta_{s_1, s_2} \delta_{t_1,t_2} 
\delta_{s_1', s_2'} \delta_{t_1',t_2'} m^{x}_{1,2;1',2'} (\bm{P}) \\
m^{x}_{1,2;1',2'} (\bm{P}) &\equiv & \nonumber 
\frac{1}{4} [\sqrt{2}]^{\delta_{n_1,0} + \delta_{n_1',0} + \delta_{n_2,0} + \delta_{n_2',0}}
\\ & & \nonumber 
\mathop{\sum_{\mu,\nu = 0}^1} 
s_{\mu}(n_1) s_{\mu}(n_2) s_{\nu}(n_1') s_{\nu}(n_2')
\\ & & \nonumber 
\quad v_{
\lambda_{\mu}(n_1), \lambda_{\mu}(n_2); \lambda_{\nu}(n_1'), \lambda_{\nu}(n_2')
} (\bm{P})
\\
M^{\Omega}_{1,2;1',2'} &=& \nonumber
\delta_{k_1,k_1'} 
\delta_{k_2,k_2'} 
\delta_{s_1,s_1'} 
\delta_{s_2,s_2'}
\delta_{t_1,t_1'} 
\delta_{t_2,t_2'}  \\ & & \nonumber
\mathop{\sum_{n_3, s_3, t_3}} \Theta(\mu - \varepsilon_3)
\\ & & \quad [ 
\delta_{n_2,n_2'} \delta_{t_1,t_3} \delta_{s_1,s_3} 
m^{d}_{n_1,n_1';n_3,n_3}(0)
\\ & & \quad \nonumber -\delta_{n_1,n_1'} \delta_{t_2,t_3} \delta_{s_2,s_3} 
m^{d}_{n_2,n_2',n_3,n_3}(0)] , 
\end{eqnarray*}
where 
$\lambda_0(n) = |n|$, $\lambda_1(n) = |n| - 1$,
$s_0(n) = 1$, and $s_1(n) = \sgn(n)$. 

The self-energy term has an interesting feature in graphene.
Note that this term involves the direct matrix elements evaluated at $P = 0$.
In the 2DEG, the conservation of internal angular momentum of the exciton requires that 
$m^{d}_{n_1,n_1';n_3,n_3}(0)$ vanish unless $n_1 = n_1'$.
Thus $M^{\Omega}$ represents a self-energy correction associated with 
single particle states which depends only on the Landau level index.
In graphene, the analogous conservation law requires only that $|n_1| = |n_1'|$.
The exchange self-energy is thus associated both with a Landau level and its 
electron-hole conjugate, with possible off-diagonal terms.  Such off-diagonal 
terms contribute to $\hat{M}$ only if the chemical potential is sufficiently low (high)
and Landau-level mixing is sufficiently strong that 
both Landau levels $-n$ and $n \ge 1$ are available for the excited electron (hole).
This circumstance never arises in our calculations, since we restrict our attention to 
$l_z \le 1$.

The interacting contribution $\Delta E^{\nu} (P) $ to the exciton dispersions
computed in this article are given in Table \ref{tab:dispersions} in terms of 
certain matrix elements, which we list below. 
For the direct contribution, 
\begin{eqnarray*}
E_3^{(0)} &\equiv& m^{d}_{1,0;1,0}  = m^{d}_{0,-1;0,-1}  
\\ &=& - \frac{1}{2}[u_{0,0;0,0}  + u_{1,0;1,0}  ]  \nonumber
\\ &=& - \frac{\beta}{8} \sqrt{\frac{\pi}{2}} \: e^{-x} [(6 + P^2) I_0(x) - P^2 I_1(x)] \\
E_3^{(1)} &\equiv& m^{d}_{2,1;2,1}  = - \frac{1}{4}[
u_{2,1;2,1} 
\\ & & \nonumber \quad 
+ u_{1,1;1,1} + u_{2,0;2,0} + u_{1,0;1,0} ] 
\\ &=& \nonumber 
-\frac{\beta}{128} \sqrt{\frac{\pi}{2}}\:e^{-x} 
[(66 + 13 P^2 + 2 P^4 + P^6) I_0(x) 
\\ & & \nonumber \quad - P^2 (23 + 4 P^2 + P^4) I_1(x)] \\
E_3^{(2)} &\equiv& m^{d}_{3,2;3,2}  = - \frac{1}{4}[
u_{3,2;3,2} 
\\ & & \nonumber \quad + u_{2,2;2,2} + u_{3,1;3,1} + u_{2,1;2,1} ]
\\ &=& \nonumber - \frac{\beta}{3072} \sqrt{\frac{\pi}{2}}\: e^{-x}
[(1398 + 237 P^2 + 10 P^4 
\\ & & \nonumber \quad \quad \quad + 43 P^6 - 6 P^8 + P^{10})I_0(x)
\\ & & \nonumber \quad - P^2 (497 + 84 P^2 + 37 P^4 - 4 P^6 + P^8) I_1(x)] \\
E_4 &\equiv& m^{d}_{0,0;0,0}  = - u_{0,0;0,0}  =  - \beta \sqrt{\frac{\pi}{2}} \: e^{-x} I_0 (x) \\
h &\equiv& m^{d}_{1,0;0,-1}  =  - \frac{1}{2} u_{1,0;1,0} 
\\ &=& \nonumber \frac{\beta}{8} \sqrt{\frac{\pi}{2}} \: e^{-x} [P^2 I_0(x) - (2 + P^2) I_1(x)],
\end{eqnarray*}
where $x = P^2 / 4$, and $I_0$ and $I_1$ are the usual modified Bessel functions.

The exchange matrix elements 
are evaluated for momentum $\bm{P} = - P \hat{y}$:
\begin{eqnarray*}
E_2^{(0)} &\equiv& m^{x}_{1,0;1,0} = m^{x}_{0,-1;0,-1} = - m^{x}_{1,0;0,-1} 
\\ & & \nonumber \quad = \frac{1}{2} v_{1,0;1,0} 
= \frac{\beta}{4} P \: e^{- 2 x} \\
E_2^{(1)} &\equiv& m^{x}_{2,1;2,1}  = \frac{1}{4}[
v_{2,1;2,1} + 2 v_{1,0;2,1} + v_{1,0;1,0} ] 
\\ &=& \nonumber 
\frac{\beta}{64} P \: e^{-2 x} [24 + 16 \sqrt{2} - 4(2 + \sqrt{2})P^2 + P^4]\\
E_2^{(2)} &\equiv& m^{x}_{3,2;3,2}  = \frac{1}{4}[
v_{3,2;3,2} + 2 v_{2,1;3,2} + v_{2,1;2,1} ]
\\ &=& \nonumber 
\frac{\beta}{1536} P e^{-2x}[
192(5 + 2 \sqrt{6}) - 96(8 + 3 \sqrt{6})P^2 
\\ & & \nonumber \quad + 8(27 + 8\sqrt{6})P^4 - 4(6+\sqrt{6})P^6 + P^8].
\end{eqnarray*}

The exchange self-energy constants diverge logarithmically
in the number of filled levels. We introduce a cutoff $-n_c$ for the index of the 
lowest filled Landau level. We define the constants
\begin{eqnarray*}
a &\equiv& \mathop{\sum_{n = -n_c}^0} [m^{d}_{1,1;n,n} (0) - m^{d}_{0,0;n,n} (0) ] \\
a^{(1)} &\equiv& \mathop{\sum_{n = -n_c}^1} [m^{d}_{2,2;n,n} (0) - m^{d}_{1,1;n,n} (0) ] \\
a^{(2)} &\equiv& \mathop{\sum_{n = -n_c}^2} [m^{d}_{3,3;n,n} (0) - m^{d}_{2,2;n,n} (0) ] \\
b &\equiv& \mathop{\sum_{n = -n_c}^{-1}} [m^{d}_{0,0;n,n} (0) - m^{d}_{-1,-1;n,n} (0) ] \\
c &\equiv& 
\mathop{\sum_{n = -n_c}^{-1}} m^{d}_{1,1;n,n} (0) - 
\mathop{\sum_{n = -n_c}^{0}} m^{d}_{0,0;n,n} (0) \\ 
d &\equiv& 
\mathop{\sum_{n = -n_c}^{-1}} m^{d}_{0,0;n,n} (0) - 
\mathop{\sum_{n = -n_c}^{0}} m^{d}_{-1,-1;n,n} (0) \\
e &\equiv& 
\mathop{\sum_{n = -n_c}^{-1}} m^{d}_{0,0;n,n} (0) -
\mathop{\sum_{n = -n_c}^{0}} m^{d}_{0,0;n,n} (0)
\end{eqnarray*}

The summation need only be carried out for $a$, $a^{(1)}$, $a^{(2)}$, and $b$ 
since 
\begin{align}
e &= - m^{d}_{0,0,0,0}(0) = \beta \sqrt{\frac{\pi}{2}} , \nonumber \\
c - a &= d - b = - m^{d}_{1,1;0,0}(0) = \frac{\beta}{4} \sqrt{\frac{\pi}{2}} .
\end{align} 
Furthermore, we find that $a - b \rightarrow 0$ in the limit $n_c \rightarrow \infty$. 
This occurs because $a$ and $b$ are exchange constants for two many-body states connected by 
the simultaneous reversal of both charge and energy, and particle-hole symmetry is obtained 
in the limit $n_c \rightarrow \infty$.  A finite cutoff results in a difference $a-b$ which 
is nonvanishing but small for $n_c \gg 1$.
However, recall that in the tight-binding approximation, there are both upper and lower cutoffs for
the kinetic energy due to the discrete nature of the lattice.  
The upper cutoff, and thus particle-hole symmetry,
is lost in passing to the continuum approximation. 
In this calculation, we assume particle-hole symmetry, and consequently $a=b$. 
We handle the cutoff by using the asymptotic form as $n_c \rightarrow \infty$:
\begin{eqnarray}
a = b &\rightarrow& 
\frac{\beta}{4\sqrt{2}}\left[ \frac{3}{2} \sqrt{\pi} 
+ \gamma + \ln (n_c) \right. 
\label{eq:ab} \\
& & \nonumber \quad \left.
+ \mathop{\sum_{n=1}^{\infty}} \left(
\frac{n^{1/2} \: \Gamma(n - 1/2)}{n!} - \frac{1}{n} \right) \right]
\end{eqnarray}
where $\gamma$ is the Euler constant.

The cutoff is determined by the fact that, due to the 4 spin and pseudospin sublevels,
there are approximately $4 n_c $ electrons per quantum of magnetic flux. For a magnetic field of 
20 T we find $n_c \approx 1870$, giving $a = b \approx 2.09$, $c = d \approx 2.41$, $a^{(1)} \approx 1.01$, 
and $a^{(2)} \approx 0.82$.
In practice, these values are rather insensitive to value of 
$n_c$ due to the logarithmic behavior in Eq. \ref{eq:ab}.

Expressions for the various branches of the energy shift 
are given in Table \ref{tab:dispersions}.  In the absence of particle-hole symmetry (as can arise from 
next-nearest neighbor hopping,) $a-b=c-d \ne 0$, and  
the expressions change somewhat; the main effect of this small difference will be a partial lifting of the 
degeneracy of $\Delta E^2$ in the cases $\gamma_0 < 4$, 
splitting it into three branches when $\gamma_0=2$ and two branches when $\gamma_0=1,3$. 

\begin{table}
\begin{tabular}{|c|c|l|}
\hline
filling & branch & dispersion \\
\hline
$\gamma_0 = 1,2,3,4$; $l_z = 1$ & $\: \Delta E^1$ & $\: a + E_3^{(0)} + 4 E_2^{(0)}$ \\
 & $\: \Delta E^2$ & $\: a + E_3^{(0)} $ \\
 & $\: \Delta E^3$ & $\: c +  E_3^{(0)} + h$ \\
 & $\: \Delta E^4$ & $\: c +  E_3^{(0)} - h$ \\
\hline
$\gamma_1 = 4$, $l_z = 1$ & $\: \Delta E^1$ & $\: a^{(1)} + E_3^{(1)}  + 4 E_2^{(1)} $ \\
 & $\: \Delta E^2$ & $\: a^{(1)} + E_3^{(1)}$ \\
\hline
$\gamma_2 = 4$, $l_z = 1$ & $\: \Delta E^1$ & $\: a^{(2)} + E_3^{(2)}  + 4 E_2^{(1)} $ \\
 & $\: \Delta E^2$ & $\: a^{(2)} + E_3^{(2)}$ \\
\hline
$\gamma_0 < 4$; $l_z = 0$ & $\: \Delta E^0$ & $\: e + E_4$ \\
\hline
\end{tabular}
\caption{Expressions for various branches of the energy shift at different filling factors.
}
\label{tab:dispersions}
\end{table}

\end{appendix}


\begin{thebibliography}{99}
\bibitem{CoulombApprox} The intervalley wavevector is of order $a^{-1}$, with $a$ the nearest-neighbor 
separation.  The lattice structure also introduces a small difference between the interaction 
potentials for electrons on the same sublattice and for those on different sublattices.  Both effects 
are suppressed by a factor $a / \ell \ll 1$ and are thus ignored for the purposes of this calculation.
\end{thebibliography}
\end{document}